\documentclass[aps,prx,reprint,superscriptaddress,floatfix,amsmath,amssymb]{revtex4-2}

\usepackage{inputenc}
\usepackage{verbatim,csquotes,bbm,mathrsfs,soul}
\usepackage{graphicx,color,mathrsfs}
\graphicspath{{./figures/}}
\usepackage{tikz}
\usepackage{circuitikz}
\usepackage[psdextra]{hyperref} 

\hypersetup{unicode=true} 
\hypersetup{colorlinks=true}
\hypersetup{bookmarksopen=true, bookmarksopenlevel=2} 

\newcommand{\avg}[1]{\langle {#1} \rangle}

\newcommand{\ket}[1]{\vert {#1}\rangle}
\newcommand{\braket}[2]{\langle{#1} \vert {#2}\rangle}
\newcommand{\ketbra}[2]{\vert {#1} \rangle \langle{#2}\vert}

\makeatletter
\newcommand\thefontsize[1]{{#1 The current font size is: \f@size pt\par}}
\makeatother

\newcommand{\tf}{t_\mathrm{f}}

\begin{document}

\title{Detecting virtual photons in ultrastrongly coupled superconducting quantum circuits}

\author{L. Giannelli}
\affiliation{CNR-IMM, UoS Universit\`a, 95123, Catania, Italy}
\affiliation{Dipartimento di Fisica e Astronomia ``Ettore Majorana'',  Universit\`a di Catania, Via S. Sofia 64, 95123, Catania, Italy}
\affiliation{INFN Sez. Catania, 95123  Catania, Italy}

\author{E. Paladino}
\affiliation{CNR-IMM, UoS Universit\`a, 95123, Catania, Italy}
\affiliation{Dipartimento di Fisica e Astronomia ``Ettore Majorana'',  Universit\`a di Catania, Via S. Sofia 64, 95123, Catania, Italy}
\affiliation{INFN Sez. Catania, 95123  Catania, Italy}

\author{M. Grajcar}
\affiliation{Department of Experimental Physics, Comenius University, SK-84248 Bratislava, Slovakia}

\author{G. S. Paraoanu}
\affiliation{QTF Centre of Excellence, Department of Applied Physics,
  Aalto University, P.O. Box 15100, FI-00076 AALTO, Finland}

\author{G. Falci}
\affiliation{CNR-IMM, UoS Universit\`a, 95123, Catania, Italy}
\affiliation{Dipartimento di Fisica e Astronomia ``Ettore Majorana'',  Universit\`a di Catania, Via S. Sofia 64, 95123, Catania, Italy}
\affiliation{INFN Sez. Catania, 95123  Catania, Italy}
\affiliation{CSFNSM, 95123 Catania, Italy}

\date{\today}

\begin{abstract}
  Light-matter interaction and understanding the fundamental physics behind, is essential for emerging quantum technologies. Solid-state devices may explore new regimes where coupling strengths are \textquote{ultrastrong}, i.e. comparable to the energies of the subsystems. New exotic phenomena occur the common root of many of them being the fact that the entangled vacuum contains virtual photons. They herald the lack of conservation of the number of excitations which is the witness of ultrastrong coupling breaking the U(1) symmetry.
  Despite more than a decade of research, the detection of ground-state virtual photons still awaits demonstration. In this work, we recognize the \textquote{conspiring} set of experimental challenges {\em and} show how to overcome them, thus providing a solution to this long-standing problem. We find that combining a superinductor-based unconventional "light fluxonium" qudit and coherent control yields a highly efficient, faithful and selective conversion of virtual photons into real ones. This enables their detection with resources available to present-day quantum technologies.
\end{abstract}

\maketitle
\section{Introduction}
\begin{figure}
  \resizebox{0.9\columnwidth}{!}{
    \includegraphics{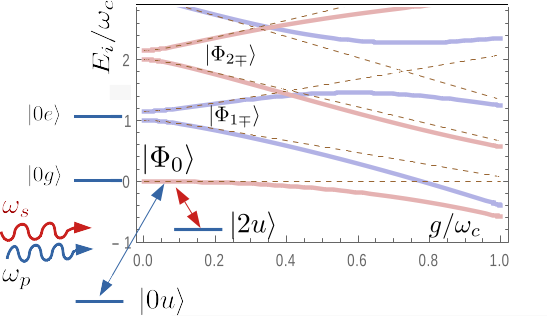}}
  \caption{Spectrum $\{E_i \}$ of the two-level Rabi model: the eigenstates $\ket{\Phi_i}$ have even (light red) and odd (light blue) parity of $\hat N$. Dashed lines show the RW approximation, i.e. the exact eigenstates of the JC Hamiltonian. If an uncoupled atomic level $\ket{u}$ is added, factorized eigenstates
    $\ket{n}\otimes \ket{u}$ also exist. The three-level "Lambda" dynamics involving transitions $\ket{0u}\!\leftrightarrow\!\ket{\Phi_0}$ (pump) and $\ket{2u}\!\leftrightarrow\!\ket{\Phi_0}$ (Stokes) reveals the entangled nature of the Rabi (false) vacuum $\ket{\Phi_0}$.}
  \label{fig:three-level}
\end{figure}

Artificial atoms (AA) and quantized modes of an electromagnetic field~\cite{ka:05-ciuti-prb-intersubbandpolariton,kr:19-forndiazsolano-rmp-usc,kr:19-kokchumnori-natrevphys-usc}
are said to be ultra-strongly coupled (USC) when the coupling strength $g$ is comparable with the natural frequencies of the uncoupled subsystems, which are the atomic energy splittings $\epsilon$ and the angular frequencies $\omega_c$ of the modes. In the last decade, the USC regime with $g$ typically $\sim 0.1-1$ times  $\omega_c$ and/or $\epsilon$ has been achieved in several different architectures of AAs~\cite{kr:19-forndiazsolano-rmp-usc,kr:19-kokchumnori-natrevphys-usc}, those based on semiconductors~\cite{ka:09-anapparabeltram-prb-ustr,ka:09-guentertredicucci-nature-switcusc,ka:12-scalari-science-USCTHz} on superconductors~\cite{ka:10-niemczyck-natphys-ultrastrong,ka:10-diazmooji-prl-blochsiegert,ka:18-magazzu-natcomms-spinboson} and on hybrid devices~\cite{22-scarlino-prx-hybridusc} being the most promising for applications. In these systems values of $g/\omega_c > 1$ have also been engineered entering the so-called deep-strong coupling regime~\cite{ka:17-yoshiharasemba-nphys-dsc,ka:17-forndiazlupascu-natphys-usctunableflux,ka:17-bayerlange-nanolett-dscsemicond}.
The simplest model of light-matter interaction, a two-level atom coupled to a single quantized harmonic mode, is the well-celebrated two-level quantum Rabi model~\cite{ka:137-rabi-pr-rabimodel,ka:11-braak-prl-rabisol}
\begin{equation}
  \label{eq:rabiH2}
  H_\mathrm{R} = \epsilon_{eg} 
  \, \ketbra{e}{e} + \omega_c  \, a^\dagger a+
  g \, (a^\dagger+a) \, \big(\ketbra{g}{e} + \ketbra{e}{g}\big) \;,
\end{equation}
where $\{\ket{g},\ket{e}\}$ are the atomic eigenstates and $a$ ($a^\dagger$) the annihilation (creation) operator of the mode acting on the space spanned by the Fock states $\{\ket{n}\}$.
The spectrum of $H_\mathrm{R}$ is shown in Fig.~\ref{fig:three-level}.
The coupling $g$ is supposed to be large enough to overcome both the mode's and the atom's decoherence rates, $g\gg \kappa, \gamma_a$. This condition marks the standard strong-coupling regime of quantum optical~\cite{kb:06-harocheraimond} and  solid-state~\cite{ka:99-imamoglu-prl-qdotJC,ka:04-wallraff-superqubit,kr:08-schoelkopf-nature-wiring,kr:20-blaisgirvingoliver-natphys-circQED} systems. When $g$ is much smaller than $\epsilon$ and $\omega_c$ the rotating wave (RW) approximation can be applied, namely only the part of the interaction conserving the number $\hat N:=  \ketbra{e}{e} + a^\dagger a$ of excitations is retained while the remaining "counter-rotating" terms are neglected. This leads to the Jaynes-Cummings (JC) Hamiltonian~\cite{JC_originalpaper,kb:06-harocheraimond} $H_\mathrm{JC}$ whose simple dynamics has been largely exploited in cavity- and circuit-QED~\cite{kr:20-haroche-natphys-cavitycircuit} for implementing quantum control~\cite{ka:03-plastinafalci-prb,ka:04-wallraff-superqubit,kr:17-nori-review-supercqed} and for many other tasks in quantum technologies~\cite{ka:17-mohsenimartinis-nature-qtech,kr:20-blaisgirvingoliver-natphys-circQED}.

In the USC regime, $g$ the counter-rotating interaction cannot be neglected breaking the conservation of $\hat N$. As a consequence, a rich non-perturbative physics is predicted to emerge, from new effects in nonlinear quantum optics to many-body physics and quantum phase transitions, with appealing applications to quantum technologies as ultrafast computation and entangled state generation~\cite{kr:19-kokchumnori-natrevphys-usc}. The hallmark of USC is the fact that the eigenstates {\em including the ground-state}, contain a significant number of (virtual) photonic and atomic excitations. Indeed while the ground-state of $H_\mathrm{JC}$ is factored in the oscillator and the atomic parts, $\ket{\phi_0} = \ket{0\,g}$, the vacuum of $H_\mathrm{R}$ Eq.(\ref{eq:rabiH2}) is entangled
\begin{equation}
  \label{eq:rabi-gs}
  \ket{\Phi_0} =\!
  \sum_{n=0}^\infty  \big(\braket{2n \,g}{\Phi_0}
  \ket{2n \,g}
  + 
  \braket{2n+1 \,e}{\Phi_0}  \ket{2n+1 \,e} \big),
\end{equation}
exhibiting a two-modes squeezed photon fields structure built on virtual photons (VPs) contained in the $n > 0$ components~\cite{ka:05-ciuti-prb-intersubbandpolariton,ka:210-ashhabnori-pra-uscdyn}. The eigenstates $\ket{\Phi_j}$ of $H_\mathrm{R}$ preserve only the parity of $\hat N$ which is even for $\ket{\Phi_0}$ (see Fig.~\ref{fig:three-level}).

It is tantalizing that the rich theoretical scenario of USC has an experimental counterpart limited so far to standard spectroscopy. What has prevented a broader experimental investigation? To gain insight into this issue we address a fundamental problem posed since the birth of the field~\cite{ka:05-ciuti-prb-intersubbandpolariton} namely the experimental detection of VPs, which still awaits demonstration despite several theoretical proposals~\cite{ka:07-deliberato-prl-uscdynamics,ka:08-takashimazeilinger-jopamat,ka:08-werlangvillasboas-pra,ka:09-deliberato-pra-extracavity,ka:15-shapirolozovik-pra,ka:12-carusotto-prl-3levusc,ka:13-stassisavasta-prl-USCSEP,ka:14-huanglaw-pra-uscraman,ka:17-distefanosavastanori-njp-stimemission,ka:17-falci-fortphys-fqmt,ka:19-falci-scirep-usc, ka:19-ridolfofalci-epj-usc, 
  ka:15-lolliciuti-prl-ancillaryphotodetection,
  ka:16-cirionori-prl-electrolumvirtphoton, ka:19-deliberato-pra-quantumvacuum,
  ka:17-cirionori-prl,
  kr:19-forndiazsolano-rmp-usc,kr:19-kokchumnori-natrevphys-usc}.
The specific question we ask is whether it is possible to overcome experimental challenges posed by available quantum hardware. This work shows that the answer is positive but not trivial. Indeed detecting VPs in an efficient and faithful way requires combining state-of-the-art technologies, such as a multilevel AA unconventional design~\cite{ka:09-manucharyan-science-fluxonium,ka:12-maslukpopdevoret-prl-superinductor,ka:19-gruenhaupt-natmaterials-superinductance,ka:19-hazard-prl-superinductances}, and a multiphoton coherent control protocol with a tailored integrated measurement technique. The implementation of this setup is feasible with present-day superconducting technology~\cite{kr:20-kjaergaardoliver-annurevcond-revsupqubits}.

While VPs are primarily a mathematical language their {\em detection} has a physical meaning since it provides a test of ground-state entanglement in the USC regime. Ground-state VPs are operatively defined by observing that
they disappear for adiabatic switching off
of the interaction $g$ but if $g$ is suddenly switched off they are released from the now uncoupled mode~\cite{ka:05-ciuti-prb-intersubbandpolariton}. They cannot be probed by standard photodetection since in the USC vacuum they are bound~\cite{ka:18-distefanonori-scirep-qphotodetection}. Therefore different methods are needed to convert VPs to detectable real excitations.
This conversion may occur in principle by modulating system parameters inducing a phenomenon similar to the dynamical Casimir effect~\cite{ka:06-ciuticarusotto-pra-inputoutput,ka:07-deliberato-prl-uscdynamics,ka:09-deliberato-pra-extracavity}. However, large values of $g$ would require faithful subnanosecond control still unavailable for present-day quantum hardware. A more viable strategy uses an AA with an additional probe-level $\ket{u}$ with lower energy and not coupled to the mode~\cite{ka:12-carusotto-prl-3levusc,ka:14-huanglaw-pra-uscraman,ka:17-falci-fortphys-fqmt,ka:17-distefanosavastanori-njp-stimemission,ka:19-falci-scirep-usc} (see Fig.~\ref{fig:three-level}). In this work, we exploit this probe-level technique aiming to provide an example of realistic experimental conditions for achieving efficient, faithful and selective detection of VPs.

The article is organized as follows. In \S\ref{sec:three-level-detection} we introduce the principles of VP-conversion to real photons and discuss experimental difficulties. In \S\ref{sec:qc-design} we study the design of a superconducting quantum circuit fulfilling all the requirements for the faithful detection of ground-state VPs. In \S\ref{sec:dynamics} we show that it is possible to detect VPs within state-of-the-art quantum technology. The result is commented in \S\ref{sec:remarks} and in the conclusions \S\ref{sec:conclusions}.
\section{Detection of virtual photons by a probe atomic level}
\label{sec:three-level-detection}
For VPs detection with a three-level AA, we consider an additional level $\ket{u}$ at lower energy than $\ket{g}$, $-\epsilon_{gu} <0$. For  illustration purposes, we assume that
the $u-g$ transition is not coupled to the mode (see Fig.~\ref{fig:three-level}) thus the
Hamiltonian reads
\begin{equation}
  \label{eq:H-3level}
  H_0 = H_\mathrm{R} -\epsilon_{gu} \,\mathbbm{1}_\mathrm{osc} \otimes \ketbra{u}{u} +  \omega_c \,a^\dagger a \otimes \ketbra{u}{u} \;.
\end{equation}
Its eigenstates are classified in two sets (see Fig.~\ref{fig:three-level}), namely the factorized states $\{\ket{nu}\}$ with eigenvalues $-\epsilon_{gu} + n \omega_c$ and the entangled eigenstates $\{\ket{\Phi_l}\}$ of the two-level $H_\mathrm{R}$, with eigenvalues $E_l$. 
The USC regime reflects in $\ket{\Phi_0}$ containing virtual excitations witnessed by the amplitudes other than $\braket{0g}{\Phi_0}$ in Eq.(\ref{eq:rabi-gs}). In particular, $\braket{2g}{\Phi_0}$ reflects the presence of a pair of VPs playing an important role in our work.

\subsection{Simple theory of virtual photon conversion}
\label{sec:vp-conversion}
The key point is that $\ket{\Phi_0}$ is a false vacuum of $H_0$ thus VPs are not bound and can be detected. An early work~\cite{ka:13-stassisavasta-prl-USCSEP} proposed to use stimulated emission pumping (SEP)~\cite{kr:01-vitanov-advatmolopt} of population $\ket{0u} \to \ket{\Phi_0}$
which is then transferred by atomic decay $\ket{g}\to \ket{u}$ to $\ket{2u}$. The process takes place only if $\braket{2g}{\Phi_0} \neq 0$ i.e. only if $\ket{\Phi_0}$ contains a pair of VPs. These are converted into two real photons in $\ket{2u}$ and can be eventually detected.
However, SEP is very inefficient~\cite{kr:98-bergmann-rmp-stirap,kr:01-vitanov-advatmolopt} since the population in $\ket{\Phi_0}$ mainly decays back to $\ket{0u}$ with no VP conversion and the conversion rate is way too small for VPs detection in relevant experimental conditions~\cite{ka:19-falci-scirep-usc} where it is estimated as $\propto |\braket{2g}{\Phi_0}|^2 \propto  (g/2\omega_c)^4$~\cite{ka:11-beaudouinblais-pra-decoherenceUSC,ka:19-falci-scirep-usc}.

To overcome this problem it has been proposed to use coherent control~\cite{ka:14-huanglaw-pra-uscraman,ka:17-falci-fortphys-fqmt} driving the AA by a two-tone classical field $W(t)=\mathscr{W}_s(t) \cos \omega_s t + \mathscr{W}_p(t) \cos \omega_p t$ which couples to the atom (see Fig.~\ref{fig:three-level}). 
We assume that the field is resonant with the two transitions $E_0 - \epsilon_u \approx \omega_p$ and $E_0 - \epsilon_u - 2 \omega_c \approx \omega_s$, all the other transitions being strongly detuned. Then standard approximations yield the $\Lambda$ driving configuration~\cite{kr:01-vitanov-advatmolopt} of
Fig.~\ref{fig:three-level} described in a rotating frame by the Hamiltonian~\cite{ka:17-falci-fortphys-fqmt}
\begin{equation}
  \label{eq:control-lambda}
  \tilde{H}_C^{\Lambda} = \frac{\Omega_p(t)}{2}
  \, \ketbra{0u}{\Phi_0} +
  \frac{\Omega_s(t)}{2}
  \, \ketbra{2u}{\Phi_0}+ \mbox{h.c.} \; .
\end{equation}
Here the Rabi amplitudes $\Omega_p(t)$ and  $\Omega_s(t)$ are proportional respectively to $\braket{0g}{\Phi_0}$ and $\braket{2g}{\Phi_0}$, besides the usual dependence on the slowly varying envelopes $\mathscr{W}_{s/p}(t)$ and on the 
AA "dipole" matrix element~\cite{kb:06-harocheraimond,ka:17-falci-fortphys-fqmt}.
This control configuration may induce deterministic population transfer
$\ket{0u} \to \ket{2u}$ by stimulated Raman adiabatic passage (STIRAP)~\cite{kr:17-vitanovbergmann-rmp,kr:01-vitanov-advatmolopt,ka:17-falci-fortphys-fqmt} 
or by Raman oscillations~\cite{ka:14-huanglaw-pra-uscraman}. Since $\Omega_{s}$ vanishes unless $\braket{2g}{\Phi_0} \neq 0$ population transfer to $\ket{2u}$ converts two VPs contained in $\ket{\Phi_0}$ to real photons. Both protocols ideally yield complete population transfer thus they {\em coherently amplify} 
up to 100\% the conversion of the aforementioned VPs.
\begin{table*}[t]
  \centering
  \begin{tabular}{|c|p{45mm}|p{55mm}|p{65mm}|}
    \hline
                                                  & {\bf Requirement}                                     & {\bf Problem}                                               & {\bf Solution}                                     \\ \hline \hline   1
                                                  & Large anharmonicity $\epsilon_{gu} \gg \epsilon_{eg}$ & Faithful VPs conversion                                     & Fluxonium-like qudit design (\S\ref{sec:remarks});
    AA biased at symmetry $Q_\mathrm{x}=\Phi_{xi}=0$
    (\S\ref{sec:case-study}).
    \\\hline  2 &
    Large splitting and mode natural frequency  $\epsilon_{eg} \sim \omega_c$ & Thermal population of the mode                        &
    AA design tradeoff (\S\ref{sec:aa-design}); large absolute energy scale $E_\mathrm{J}$ (\S\ref{sec:remarks})
    \\\hline  3 &
    Not too large probe-splitting $\epsilon_{gu}$ & Reliable microwave control                            &
    AA design tradeoff and not too large $E_\mathrm{J}$
    \\\hline  4 &
    Large $\gamma_{ge}/\epsilon_{eg}$             & Attaining USC regime                                  & Galvanic coupling with superinductors (\S\ref{sec:remarks})
    and/or design of not too large reference
    $\epsilon_{eg}$                                                                                                                                                                                                          \\\hline  5 &
    Small $\gamma_{gu} \ll \gamma_{eg}$           &
    Faithful VPs conversion
                                                  & AA design tradeoff
    \\\hline  6 &
    Large enough $q_{gu} \propto \gamma_{gu}$     & Large  enough
    coupling to the external control field        &
    Small $C_1$ $\to$  "light" fluxonium qudit $E_{C_1} \sim E_\mathrm{J}$ preventing localization the side minima
    \\\hline
  \end{tabular}
  \caption{Spectral requirements for the AA in a  superconducting-based USC quantum system allowing faithful detection of ground-state VPs by the probe-level technique. We also need a large enough absolute energy scale $E_\mathrm{J}/(2 \pi) \gtrsim 10$GHz limiting the thermal population of the mode when the subsystems are nearly resonant,  $\epsilon_{eg} \approx \omega_c$ (see Fig.~\ref{fig:parametric}b), and a small enough ”mass” $C_1$ preventing trapping into the side minima of the potential. Conditions (1,2,3) are  conflicting and require looking for a tradeoff with circuit parameters (see Fig.~\ref{fig:parametric}b,d). Conditions (2,4) are conflicting as well as conditions (4,5,6) and require looking for a design tradeoff better achieved if the AA is driven via the charge port  (see Fig.~\ref{fig:parametric}c,e).}
  \label{tab:requirements-spectral}
\end{table*}

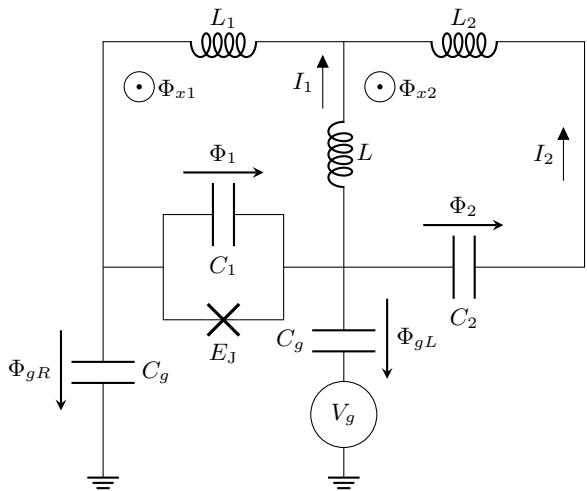
\begin{figure}
  \makeatletter
\def\TikzBipolePath#1#2{\pgf@circ@bipole@path{#1}{#2}}
\makeatother
\newlength{\ResUp}
\newlength{\ResDown}
\newlength{\ResLeft}
\newlength{\ResRight}

\ctikzset{bipoles/josephsonjunction/height/.initial=.30}   
\ctikzset{bipoles/josephsonjunction/width/.initial=.30}    
\pgfcircdeclarebipole{}                                    
{\ctikzvalof{bipoles/josephsonjunction/height}}
{josephsonjunction}                                        
{\ctikzvalof{bipoles/josephsonjunction/height}}
{\ctikzvalof{bipoles/josephsonjunction/width}}
{                                                          
  \pgfextracty{\ResUp}{\northeast}                         
  \pgfextracty{\ResDown}{\southwest}
  \pgfextractx{\ResLeft}{\southwest}
  \pgfextractx{\ResRight}{\northeast}
  \pgfsetlinewidth{3\pgfstartlinewidth}
  \pgfmoveto{\pgfpoint{\ResLeft}{\ResDown}}
  \pgflineto{\pgfpoint{\ResRight}{\ResUp}}
  \pgfmoveto{\pgfpoint{\ResRight}{\ResDown}}
  \pgflineto{\pgfpoint{\ResLeft}{\ResUp}}
  \pgfusepath{draw}
  \pgfsetlinewidth{\pgfstartlinewidth}
  \pgfmoveto{\pgfpoint{\ResLeft}{0}}
  \pgflineto{\pgfpoint{\ResRight}{0}}
  \pgfusepath{draw}
}
\def\circlepath#1{\TikzBipolePath{josephsonjunction}{#1}}
\tikzset{josephsonjunction/.style = {\circuitikzbasekey, /tikz/to path=\circlepath, l=#1}}

\newcommand\mySQUID[4]{
  \draw #2 to[josephsonjunction, l_=#3] ++
  (\hlen/2,0) to ++ (0,\vlenshort/2) coordinate(#1-out) to ++
  (0,\vlenshort/2) to[C, l=#4] ++ (-\hlen/2,0) to ++
  (0,-\vlenshort/2) coordinate(#1-in) to ++ (0,-\vlenshort/2);
}

\newcommand\myFLUX[2]{
  \draw #1 circle [radius=0.2] node[right,
  shift={(0.13cm,-0.03cm)}] {#2};
  \fill #1 circle[radius=1pt];
}

\newcommand\myVAR[3]{
  \draw[-stealth, thick] #1 -- node[above] {#3} ++ #2;
}

\def\hlen{3.2}
\def\vlen{3}
\def\vlenshort{1.4}

\begin{circuitikz}
  \mySQUID{squid1}{(\hlen/4,-\vlenshort/2)}{$E_{\mathrm{J}}$}{$C_1$}
  \draw (0,0) to (squid1-in) (squid1-out) to (\hlen,0);
  \draw (\hlen,0) to[L, l_=$L$,  f>=$I_1$] (\hlen,\vlen) to[L, l_=$L_1$]
  (0,\vlen) to (0,0);
  
  \myFLUX{(0.15*\hlen,0.8*\vlen)}{$\Phi_{x1}$}
  \myFLUX{(1.15*\hlen,0.8*\vlen)}{$\Phi_{x2}$}

  \myVAR{(\hlen/3,0.9*\vlenshort)}{(\hlen/3,0)}{$\Phi_1$}
  \myVAR{(1.33*\hlen,0.4*\vlenshort)}{(\hlen/3,0)}{$\Phi_2$}

  \draw (\hlen,0) to[C, l_=$C_2$]  ++ (\hlen,0) to[short, f>^=$I_2$] ++
  (0,\vlen) to[L, l_=$L_2$] ++ (-\hlen,0);

  \draw (0,0) to[C=$C_g$] (0,-2*\vlenshort) node[tlground]{};
  \draw[-stealth, thick] (-0.4*\vlenshort, -0.6*\vlenshort) -- node[left]
  {$\Phi_{gR}$} ++ (0,-\hlen/3);


  \draw (\hlen, 0) -- (\hlen,-0.4*\vlenshort) to[C, l_=$C_g$] ++ (0,-0.6*\vlenshort)
  node[circle,draw,pos=3.5,minimum size=25pt,fill=white] {$V_g$} -- ++ (0,-\vlenshort) node[tlground]{}; 
  
  \draw[-stealth, thick] (1.18*\hlen, -0.3*\vlenshort) -- node[right]
  {$\Phi_{gL}$} ++ (0,-\hlen/3);
\end{circuitikz}
  \caption{(a) Equivalent circuit of a two-loop superconducting device. Loop 1 containing a Josephson junction implements the AA, and loop 2 plays the role of the mode. The system is 
  driven by a gate voltage $V_g(t)$ acting on the AA's charge ($q$-port) and possibly biased by external magnetic fluxes $\Phi_{xi}$ concatenated with each loop $i=1,2$ ($\gamma$-port).}
  \label{fig:equivalent-circuit}
\end{figure}

\subsection{Experimental challenges}
\label{sec:exp-challenges}
The hardware for VP detection must meet
several requirements.
First, coherent amplification requires good coherence properties. From this point of view, superconducting quantum hardware is a more promising platform than all-semiconductor systems which have  
 poor coherence properties in the USC regime~\cite{kr:19-forndiazsolano-rmp-usc}. Also, advanced multilevel control at microwave frequencies has been successfully implemented in superconducting quantum devices~\cite{kr:20-kjaergaardoliver-annurevcond-revsupqubits,ka:16-kumarparaoanu-natcomm-stirap,ka:16-xuhanzhao-natcomm-ladderstirap,ka:16-vepsalainen-photonics-squtrit}. Moreover, efficient detection of few excess photons in the harmonic mode requires  measurement schemes more sophisticated than standard spectroscopy routinely used in USC experiments. Again the superconductor quantum technology developed in the last decades allows photodetection at microwave frequencies~\cite{kr:17-nori-review-supercqed} with resolution down to the single-photon level~\cite{ka:11-eichlerwallraff-prl-photontomography,ka:21-curtisschoelkopf-pra-pnrdetection}.
Therefore in this work, we will exploit all-superconducting USC quantum hardware.

The major problem with superconducting AAs is implementing a three-level system with $\ket{u}$ sufficiently uncoupled to the mode. Indeed a stray coupling $g_{ug}\neq 0$ between the mode and the u-g transition opens a new channel for photon-pair production even in the absence of counter-rotating interaction terms interaction~\cite{ka:19-falci-scirep-usc}. Therefore photon conversion may be unfaithful i.e. detecting photons at the end of the protocol is not always a "smoking gun" of the existence of ground-state VPs.

Faithful VP conversion requires large anharmonicity of the AA, $\epsilon_{gu} \gg \omega_c \approx \epsilon_{eg}$, and a very small stray coupling $g_{ug} \ll g$ (see Tab.~\ref{tab:requirements-spectral}). Unfortunately, these conditions are not met in standard superconducting hardware as the transmon or the flux-qubit design~\cite{ka:19-falci-scirep-usc} where strategies based on the $\Lambda$ configuration as proposed in Refs.~\cite{ka:13-stassisavasta-prl-USCSEP,ka:14-huanglaw-pra-uscraman,ka:17-falci-fortphys-fqmt,ka:17-distefanosavastanori-njp-stimemission} fail.

\section{Design of the quantum circuit}
\label{sec:qc-design}
The aim of this work is to provide an example of a quantum circuit and a protocol allowing demonstration of the faithful detection of VPs in the ground state of a USC system by available quantum resources. To this end, we look for a selected case study postponing to future work a systematic optimization of the problem. The simplest option is to model the multilevel AA by a superconducting quantum interference device (SQUID) coupled galvanically to an LC oscillating circuit by a large inductance $L$. This device corresponds to the two-loop equivalent circuit shown in Fig.~\ref{fig:equivalent-circuit}. We first look for a design yielding the desired spectral properties, summarized in Table~\ref{tab:requirements-spectral}.

In a mechanical analogy, the circuit of Fig.~\ref{fig:equivalent-circuit} is equivalent to a pair of one-dimensional fictitious particles moving in a potential. We choose as coordinates the flux variables~\cite{ka:17-VoolDevoret-IJCTA-circuitqed} $\Phi_i$ attached to the   capacitor in loop $i=1,2$ and a combination $\Phi_g$ of the flux variables attached to the ground capacitors.
The electrostatic energy stored in the capacitors yields the \textquote{kinetic energy} while the potential energy is determined by the Josephson tunnelling (energy $E_\mathrm{J}$) and by the inductive energy which is a bilinear form in the $\Phi_i$s. The Hamiltonian of the quantum circuit reads
(see Appendix~\ref{sec:qcircuits})
\begin{equation}
  \label{eq:Hqc}
  \begin{aligned}
    H_\mathrm{qc} = \sum_{i=1,2} & \frac{\hat Q_i^2}{2 C_i} + \frac{Q_\mathrm{x}}{C_1} \, \hat Q_1- E_\mathrm{J} \, \cos \frac{2 \pi \hat \Phi_1}{\Phi_0}
    \\
                          & + \frac{1}{2} \sum_{ij} (\hat \Phi_i + \Phi_{xi}) \, [\mathbbm{L}^{-1}]_{ij}\,
    (\hat \Phi_j + \Phi_{xj}) \; ,
  \end{aligned}
\end{equation}
where $[\hat \Phi_i, \hat Q_i]= i \hbar$ are conjugated variables and $\Phi_0 := h/2e$ is the flux quantum.
The Hamiltonian depends parametrically on the bias charge $Q_\mathrm{x}=C_g V_g/2$ and on the fluxes of the magnetic fields piercing the SQUID ($\Phi_{x1}$) and the LC loop ($\Phi_{x2}$) which are used to bias and to drive the system. In Eq.(\ref{eq:Hqc}) we made the usual assumption $C_g \ll C_1$. For the circuit in Fig.~\ref{fig:equivalent-circuit} the inductance matrix is given by
\begin{equation}
  \label{eq:inductance-matrix}
  \mathbbm{L} =
  \left(\hskip-3pt
  \begin{array}{cc}
      L_1 + L & - L
      \\
      -L      & L_2 + L
    \end{array}\hskip-3pt
  \right) \; ,
\end{equation}
where the mutual inductance $M$ has been neglected with respect to the galvanic coupling, $M\ll L$.

We stress that $H_\mathrm{qc}$ provides an effective Hamiltonian describing more general circuits than the one in Fig.~\ref{fig:equivalent-circuit}. For instance, it also models the relevant dynamics of a multi-junction AA or a transmission-line resonator, and
$L$ may describe a Josephson array-based super-inductor~\cite{ka:09-manucharyan-science-fluxonium,ka:12-maslukpopdevoret-prl-superinductor,ka:19-gruenhaupt-natmaterials-superinductance,ka:19-hazard-prl-superinductances}. Therefore our investigation covers a wide class of devices.

We treat the case study where all the external bias parameters  $(Q_\mathrm{x},\Phi_{xi})$ vanish
and we start with the time-independent problem.
We split $H_\mathrm{qc} = H_\mathrm{AA}(\Phi_1, Q_1) + V(\Phi_1,\Phi_2)+ H_\mathrm{LC}(\Phi_2, Q_2)$ in parts referring to the AA, to the interaction and to the LC mode respectively.
We take the Josephson energy $E_\mathrm{J}$ of the SQUID as the reference energy scale and define the other relevant scales as
\begin{equation}
  \label{eq:enscalesAA}
  E_{C_i}= \frac{2 e^2}{ C_i} \qquad ; \qquad
  \mathbbm{U} :=  \Big(\frac{\hbar}{2 e} \Big)^2 \mathbbm{L}^{-1}
  \; ,
\end{equation}
$E_{C_i}$ being the Cooper-pair's charging energies for each loop of the circuit while $\mathbbm{U}$ is the matrix of the inductive energies. These scales parametrize the Hamiltonian in dimensionless variables. Introducing the gauge-invariant Josephson phase $\hat \gamma := 2 \pi \hat \Phi_1/\Phi_0$ and the reduced charge $\hat q = \hat Q_1 / (2e)$ of the AA we obtain
\begin{equation}
  \label{eq:quantHAA}
  \hat H_\mathrm{AA} = E_C \,\hat q^2 - E_\mathrm{J} \,\cos \hat \gamma + \frac{1}{2}
  U_{11} \,\hat \gamma^2
  \; ,
\end{equation}
where $[\hat \gamma, \hat q]=i$.
Then we represent $(\hat \Phi_2,\hat Q_2) $ by the ladder operators of the mode and we write
\begin{equation}\label{eq:quantHosc}
  \hat H_\mathrm{LC}  = \sqrt{2 E_{C_2} U_{22}}  \, a^\dagger a
  \; ,
\end{equation}
which identifies $\omega_c = \sqrt{2 E_{C_2} U_{22}}$. Finally
\begin{equation}
  \label{eq:quantHint}
  \hat V = U_{12}  \, \hat{\gamma} \,
  \frac{2 \pi \hat \Phi_2}{ \Phi_0}
  =
  U_{12}  \,
  \Big(\frac{E_{C_2}}{2 U_{22}}\Big)^\frac{1}{4}\,
  \hat{\gamma} \,(a^\dagger + a)
  \; 
\end{equation}
is the interaction between the mode and the AA whose phase $\hat \gamma$  plays the role of the dipole operator (cf. \S\ref{sec:vp-conversion}).

\begin{figure}[t!]
  \centering
  \includegraphics[width=\linewidth]{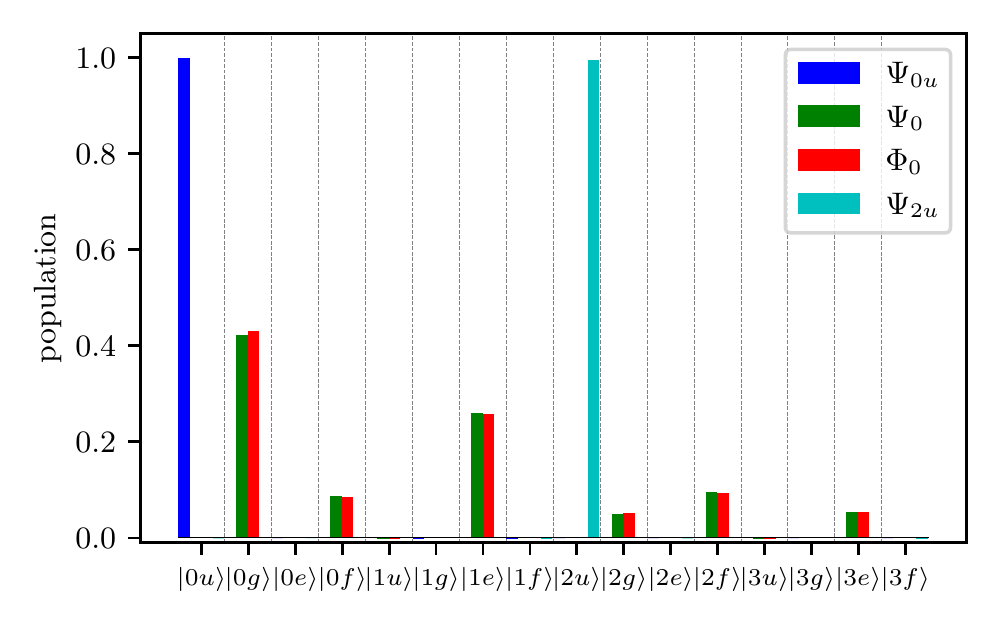}
  \caption{Decomposition in the factorized basis
    of the eigenstates of $H_\mathrm{qp}$ relevant in population transfer,
    for the set of parameters 2 in Tab.~\ref{tab:circuit-parameters}. The initial (blue) and the target (cyan) states are almost factorized $\ket{\Psi_\mathrm{nu}}\approx \ket{nu}$. On the other hand, the (false) Rabi ground state $\ket{\Psi_0}$  (full green) is similar to the  vacuum $\ket{\Phi_0}$ of the EQR model (red). Both  show a large multilevel entanglement implying that many atomic states must be considered in the problem.}
  \label{fig:decomposition}
\end{figure}

Introducing eigenvalues $\epsilon_i$  and eigenvectors $\{\ket{i},\,i=0, \dots, \infty\}$ of $H_\mathrm{AA}$ we finally
cast the quantum circuit described by $H_\mathrm{qc}$ into an extended quantum Rabi (EQR) model, the mode being coupled to a multilevel AA with energy splittings $\epsilon_{ji}:=\epsilon_j-\epsilon_i$.
The Hamiltonians in \S\ref{sec:three-level-detection} are obtained by truncating the AA to the three lowest levels $i=u,g,e$ and identifying
\begin{equation}
  \label{eq:g-eta}
  g = U_{12}  \,
  \Big(\frac{E_{C_2}}{2 U_{22}}\Big)^\frac{1}{4}\, \gamma_{ge}
  \quad ; \quad
  g_{ug} = \frac{\gamma_{ug}}{\gamma_{ge}}\, g \;.
\end{equation}
We denote by $(\ket{\Psi_m},E_m)$ eigenvectors and eigenvalues of the EQR model $H_\mathrm{qc}$. Since we are interested in a regime where the atomic $\ket{u}$ is weakly coupled to the mode we can still use the same quantum numbers of the "uncoupled level" limit. In particular we define $\ket{\Psi_{nu}}$ which reduce to $\ket{nu}$ when $g_{ug} \to 0$. A subset of the other $\ket{\Psi_m}$ reduces to the eigenstates $\ket{\Phi_j}$ of the two-level Rabi model  when all couplings but $\gamma_{ge}$ vanish. The false vacuum is given by $\ket{\Psi_0} = \sum_{n, i=0}^\infty \frac{1 }{ 2} \big[1 - (-1)^{n+i}\big] \ket{n i} \braket{n i}{\Psi_0}$ (see Fig.~\ref{fig:decomposition}). The number of excitations is redefined as $\hat N= a^\dagger a+ \sum_{i=0}^\infty (i-1) \ketbra{i}{i}$ and its parity is conserved also in the EQR model. The ideal coherent protocols are expected to yield complete population transfer $\ket{\Psi_{0u}} \to \ket{\Psi_{2u}}$ via the intermediate state $\ket{\Psi_{0}}$ which is never populated.

\subsection{Design in a case study}
\label{sec:case-study}
\begin{figure}[b]
  \centering
  \includegraphics{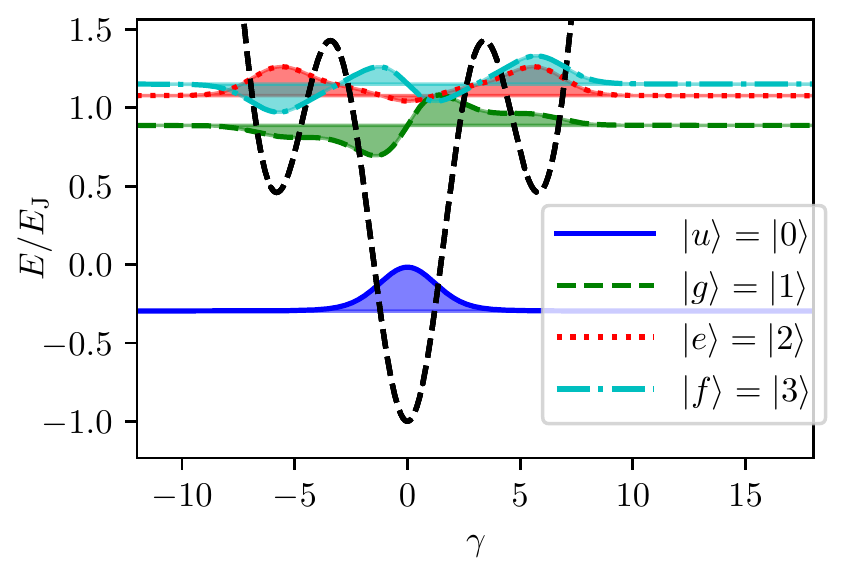}
  \caption{Potential landscape, spectrum and wavefunctions of the AA for $Q_\mathrm{x}=\Phi_{xi}=0$ and for the set of parameters (2) in  Table~\ref{tab:circuit-parameters}. The mode couples resonantly with the AA levels $\ket{g}$ and $\ket{e}$ the overlap between the wavefunctions $\braket{\gamma}{g}$ and $\braket{\gamma}{e}$ yielding to a sufficiently large dipole matrix element. The AA's ground state $\ket{u}$ must be sufficiently but not totally decoupled thus also the overlap  between $\braket{\gamma}{u}$ and $\braket{\gamma}{g}$ must be non-vanishing. Notice finally that higher-energy levels as $\ket{f}$ can be also coupled to the mode.}
  \label{fig:artificial-atom}
\end{figure}

Our choice for the case-study bias parameters is suggested 
by the shape of the AA potential in Eq.(\ref{eq:quantHAA}).
It is shown in Fig.~\ref{fig:artificial-atom} for a parametrization of interest, presenting a single absolute minimum and side relative minima (Fig.~\ref{fig:artificial-atom}) thus it is likely to implement the anharmonicity requirement $\epsilon_{gu} \gg \epsilon_{eg}$ (see Table~\ref{tab:requirements-spectral}). Moreover, for $Q_\mathrm{x}=\Phi_{xi}=0$ the potential has robust charge and flux symmetries which minimize the AA decoherence due to low-frequency noise~\cite{kr:14-paladino-rmp} and enforce \textquote{parity} selection rules for both operators $\hat \gamma$ and $\hat q$. Thus  when driving the system unwanted transitions, as those between states $u-e$, are suppressed and the problem is simplified.
\begin{figure*}[t]
  \centering
  \includegraphics{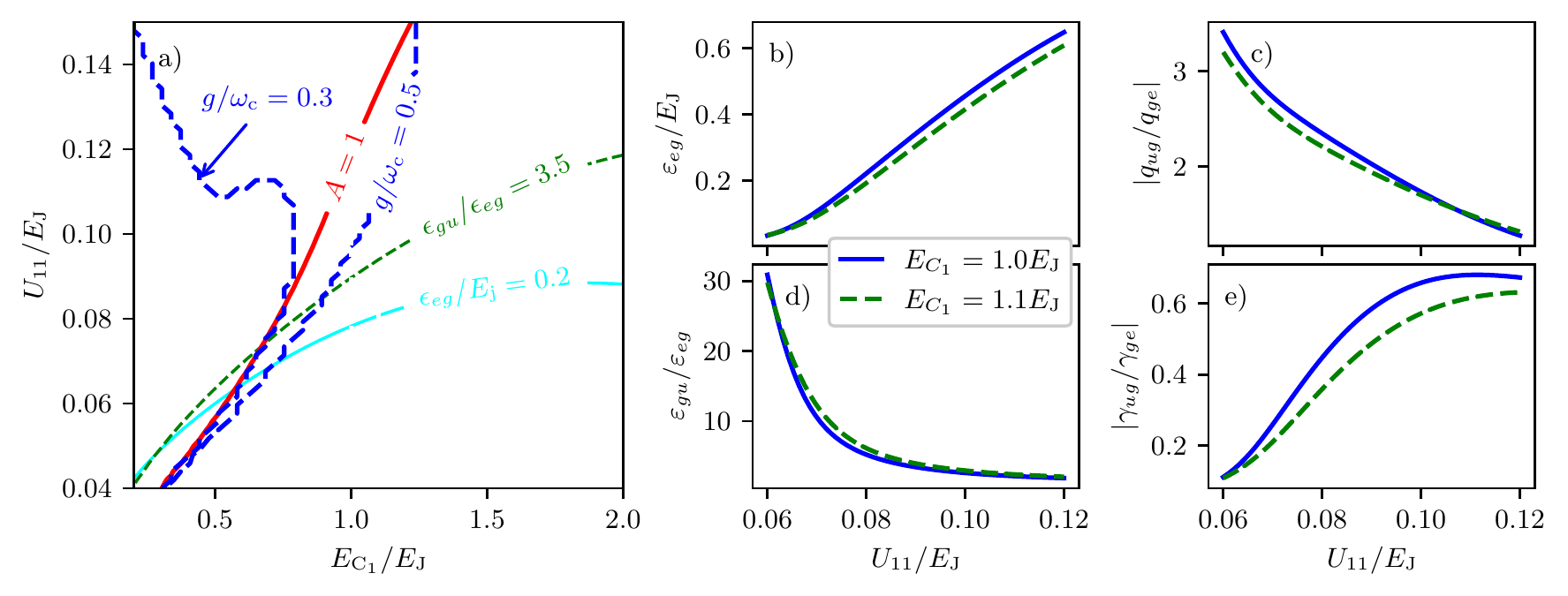}
  \caption{Spectral properties of $H_\mathrm{AA}$ as a function of the AA parameters $E_{C_1}$ and $U_{11}$, in units of $E_\mathrm{J}$. (a)
  Partition of the space of parameters according to the spectral requirements on the splitting $\epsilon_{eg}/E_\mathrm{J}$ and the probe splitting $\epsilon_{gu}/\epsilon_{eg}$ (the green and cyan arrows point towards increasing values), the  criterion Eq.(\ref{eq:figure-merit-simple}) for faithful conversion (region 
  to the right of the red curve); in the regions over the blue lines, no acceptable solutions for the whole quantum circuit with the $g/\omega_c$ exist. The interesting case studies of table~\ref{tab:circuit-parameters} (circles) lie in a narrow region of the space of parameters.
  Right panels show spectral quantities in this region (indicated by the vertical grey lines in the subfigure (a)),  on the lines $E_{C_1}/E_\mathrm{J} = 1.0$ (blue solid lines) and $E_{C_1}/E_\mathrm{J} = 1.1$ (dashed lines): (b) the coupling splitting $\epsilon_{eg}/E_\mathrm{J}$; (c) the ratio $\epsilon_{gu} /\epsilon_{eg}$, the ratio of the matrix elements (d) $\gamma_{ug}/\gamma_{ge}$ entering the coupling of the AA with the mode ;  (e) $q_{ug}/q_{ge}$ entering the coupling of the AA to the external control field. The circles represent the case studies of Table~\ref{tab:physical-param}, the arrows indicate the trend of the investigated quantities towards favourable figures.}
  \label{fig:parametric}
\end{figure*}

We observe that the system's design must satisfy several spectral requirements which are often conflicting. For instance, faithful VP conversion requires a large AA anharmonicity, $\epsilon_{gu} \gg \epsilon_{eg}$. At the same time, we need a large enough $\epsilon_{eg} \sim \omega_c$ to limit the thermal population of the mode. In order to achieve faithful conversion, matrix elements must be such that ${\gamma}_{ge} \gg {\gamma}_{ug}$. At the same time, we need ${\gamma}_{ug}$ to be large enough to allow effective coherent driving.
To this end, the \textquote{particle} $\hat \Phi_1$ should not be \textquote{trapped} in the side minima of the potential of Fig.~\ref{fig:artificial-atom} thus its \textquote{mass} $C_1$ must be sufficiently small.

In Tab.~\ref{tab:requirements-spectral} we summarize the spectral requirements. In what follows, we look for a set of energy scales $E_\mathrm{J}$, $E_{C_i}$ and $\mathbbm{U}$ allowing us to achieve a positive tradeoff between conflicting requirements. We anticipate that a suitable parametrization can be found (see Fig.~\ref{fig:artificial-atom} and Table~\ref{tab:circuit-parameters}) but at least a fourth atomic level $\ket{f}$ must enter the game.

\subsubsection{Design of the AA}
\label{sec:aa-design}
We focus on $H_\mathrm{AA}$ and determine the spectrum and the matrix elements $\gamma_{ij}$ as functions of $(E_\mathrm{J}/E_{C_1},U_{11}/E_\mathrm{J})$.
Keeping in mind that we need large enough $\epsilon_{gu}$ (Fig.~\ref{fig:parametric}d) and not too small $\gamma_{ug}$  (Fig.~\ref{fig:parametric}e) the region of interest is restricted by the quest that $\ket{u}$ is \textquote{sufficiently} decoupled to ensure faithful VPs conversion.  Good candidates are AAs such that
\begin{equation}
  \label{eq:figure-merit-simple}
  A = \frac{\epsilon_{gu} - \epsilon_{eg} }{ 2
    \epsilon_{eg}} \, \frac{\gamma_{ge}^2 }{ \gamma_{ug}^2} \gg 1   \; . \end{equation}
This criterion is obtained by asking that at resonance, $\omega_c = \epsilon_{eg}$, the effective second-order USC coupling necessary for VPs conversion, $g^2 / (2 \omega_c)$, overwhelms the RW stray coupling  $g^2_{ug}/(\epsilon_{gu}-\omega_c)$ responsible for the unwanted output photons.
It indicates that the region of interest must lie to the right of the line $A=1$ in Fig.~\ref{fig:parametric}a. Accurate figures of merit for faithful VPs conversion involve the whole coupled system and will be derived in \S\ref{sec:connection-dyn}.

The relevant region is then restricted by asking that $\epsilon_{gu}/\epsilon_{eg}$ is sufficiently large and $\epsilon_{eg}/E_\mathrm{J}$ is not very small which roughly happens between the green and the cyan line in Fig.~\ref{fig:parametric}a.
Interesting case studies are reported in Tab.~\ref{tab:requirements-spectral} and depicted by 
the open circles in Fig.~\ref{fig:parametric}a.
Focusing on this region we study  $\epsilon_{gu}/\epsilon_{eg}$ and $\epsilon_{eg}/E_\mathrm{J}$ in Fig.~\ref{fig:parametric}b,c. The red arrows in the figures mark the favourable trend for these two quantities making clear that meeting the requirements for both demands for a tradeoff. Moreover, even if the favourable region in Fig.~\ref{fig:parametric}a seems to extend for increasing $E_C/E_\mathrm{J} $ a further restriction prevents exploring larger values. Indeed, since  $\omega_c \sim \epsilon_{eg}$ scales with the reference energy $E_\mathrm{J}$ (see Fig.~\ref{fig:parametric}b) this latter has to be large enough, say $E_\mathrm{J}/(2 \pi) \sim 10-20\,$GHz, to limit the thermal population of the mode. At the same time, as argued at the beginning of this section, a small enough "mass" $C_1$ is needed to achieve a sufficient coupling of the external drives. Both requirements are not easily met in a Josephson junction therefore we seek the largest possible $C_1$ restricting our search to a region near $E_C/E_\mathrm{J} \sim 1$. This is how we selected the sample points in Tab.~\ref{tab:circuit-parameters} as the case studies for our investigation.

\subsubsection{Design of the coupled system}
We now study the design of the whole coupled system. We fix $\omega_c /\epsilon_{eg}$  and $g/\omega_c$ so we can determine  for each $(E_\mathrm{J}/E_{C_1},U_{11}/E_\mathrm{J})$ the remaining energy scales by inverting the equations $\hbar \omega_c =\sqrt{2 E_{C_2} U_{22}}$ and the first Eq.(\ref{eq:g-eta}) this leaving one undetermined parameter. For instance, we can set $L_1/L =0$ which is a physically acceptable choice having in mind a design where $L \gg L_1$. This simplifies the analysis since
$$
  \mathbbm{U} =
  \frac{(\hbar/2 e)^2 }{ L L_2}
  \left(\hskip-3pt \begin{array}{cc}
      L+L_2 & L
      \\
      L     & L
    \end{array}\hskip-3pt\right)
$$
yielding $U_{12}= U_{22}$.
Results for $\omega_c /\epsilon_{eg}=1$ are shown in Fig.~\ref{fig:parametric} for  the lines $E_{C_1}/E_\mathrm{J}=1.0,\, 1.1.$ and reported in Tab.~\ref{tab:circuit-parameters} in terms of $E_\mathrm{J}/E_{C_2}$ and $L/L_J = E_\mathrm{J}/U_{22}$,
where $L_J:=\hbar^2/(2 e)^2 \times 1/E_\mathrm{J}$ is the Josephson inductance. Once the energies are found, circuit parameters are determined by inverting Eqs.~(\ref{eq:enscalesAA}). Implications for the implementation of the device will be discussed in \S\ref{sec:remarks}.

Notice that not always this procedure yields an acceptable solution. Indeed there are regions of the
$(E_{C_1}, U_{11})$ plane where we find an unphysical not positively defined inductance matrix. Results in Fig.~\ref{fig:parametric} suggest that the acceptable space of parameters shrinks for increasing $g$, limiting to $g/\omega_c \lesssim 0.5$ the useful region for investigating VPs conversion. Remarkably, we are already well inside the non-perturbative USC region~\cite{kr:19-forndiazsolano-rmp-usc} allowing us to fully explore the new effects in this regime.
\begin{table*}[t!]
  \begin{tabular}{|c||c|c|c||c|c|c|c|c|c|c|c|}
    \hline \rule[-7pt]{0pt}{17pt}
         & $\frac{g }{ \omega_c}$ & $\frac{E_\mathrm{J} }{ E_{C_1}}$ & $\frac{U_{11} }{ E_\mathrm{J}}$ & $\frac{\epsilon_{eg} }{ E_\mathrm{J}}$ & $\frac{E_\mathrm{J} }{ E_{C_2}}$ & $\frac{L }{ L_J}$ & $\frac{L_2 }{ L_J}$ & $\frac{g_{ef} }{ \epsilon_{fe}}$ & $A_q$ & $A_q^\prime$ & $A_\gamma$ \\ \hline 1 &
    0.5  & 0.9                    & 0.08                    & 0.19                   & 0.82                          & 15.3                    & 67.6              & 12.7                & 186                              & 115   & 51                        \\ \hline 2&
    0.5  & 0.9                    & 0.081                   & 0.20                   & 0.83                          & 15.6                    & 59.7              & 13.4                & 167                              & 102   & 53                        \\ \hline 3&
    0.5  & 0.9                    & 0.083                   & 0.22                   & 0.85                          & 16.2                    & 47.3              & 14.7                & 138                              & 75    & 61                        \\ \hline 4 &
    0.5  & 0.9                    & 0.087                   & 0.267                  & 0.89                          & 18                      & 31.6              & 16.8                & 89                               & 49    & 77                        \\ \hline 5 &
    0.5  & 1.0                    & 0.08                    & 0.22                   & 1.16                          & 19.4                    & 35.3              & 26                  & 101                              & 274   & JC ph.                    \\ \hline  6  &
    0.38 & 1.1                    & 0.083                   & 0.29                   & 0.99                          & 24.9                    & 23.4              & 40.8                & 34                               & 3.9   & JC ph.                    \\ \hline
  \end{tabular}
  \caption{\label{tab:circuit-parameters} Circuit parameters for $L_1=0$,  $\omega_c=\epsilon_{eg}$ and $g/\omega_c=0.38,0.5$, for selected points of the plot of Fig.~\ref{fig:parametric}a. Also reported the figures of merit for faithful VPs conversion $A(\hat q)$ Eq.~\ref{eq:figmerit-Aq}) and $A^\prime(\hat q)$ Eq.(\ref{eq:figmerit-Apq}) for the $q$-port and the analogous $A(\hat \gamma)$ for the $\gamma$-port (see \S~\ref{sec:driving-options}) obtained by substituting $\hat q \to \hat \gamma$ in Eq(\ref{eq:figmerit-Aq}). The protocol is nearly ideal for samples 1-3 while for samples 4-6 extra photons are produced by the JC channel which can be discriminated by post-selection (see \S\ref{sec:remarks}).}
\end{table*}

We finally remark that for the parameters we selected levels of the AA with energy larger than $\epsilon_e$ are also coupled non-perturbatively to the mode as witnessed by the values of $g_{ef}/\epsilon_{fe}$ in Tab.~\ref{tab:circuit-parameters}
(see also Fig.~\ref{fig:decomposition}) thus the quantum circuit described by $H_\mathrm{qc}$ implements an EQR model.
We will prove in \S\ref{sec:connection-dyn} that this has no consequence at the fundamental level but the quantitative impact is not negligible thus reliable conclusions on the significance of experimental results require taking into account many levels of the AA.

\subsection{Driven quantum circuit}
\label{sec:connection-dyn}
We now consider driving the quantum circuit of Fig.~\ref{fig:equivalent-circuit} by a voltage $V_g(t)$. The drive enters Eq.(\ref{eq:Hqc}) via a term proportional to $Q_\mathrm{x}(t)$ thus it couples via the \textquote{$q$-port} of the AA Hamiltonian Eq.(\ref{eq:quantHAA}) and  it 
is described by adding to it the control term
\begin{equation}
  \label{eq:control-q}
  H_\mathrm{C}(t) = W(t) \, \hat q \; ,
\end{equation}
where $W(t) =e C_g V_g(t)/C_1$. To justify this choice we observe that since $[\hat \gamma, H_\mathrm{AA}] = 2 i \,E_{C_1}\, \hat q$ matrix elements  of $\hat q$ and $\hat \gamma$ in the AA eigenbasis are related
\begin{equation}
  \label{eq:q-gamma}
  \braket{i|\hat q}{j} = i \, \frac{\epsilon_{ij}}{ 2 E_C} \braket{i|\hat \gamma}{j} \; .
\end{equation}
Thus by increasing the anharmonicity $\epsilon_{gu}/\epsilon_{eg}$ of the spectrum 
the ratio between the coupling of the drive to the $u-g$ transition ($q_{ug}$) and the stray coupling $q_{ge}$ 
increases. {\em At the same time} the ratio between the stray coupling to the mode ($\gamma_{ug}$) and the USC coupling to the $g-e$ transition ($\gamma_{ge}$) increases.
Figs.~\ref{fig:parametric}d,e shows this for the region of parameters we selected, all the AA matrix elements following the favourable trend (direction of the red arrows) if $U_{11}/E_\mathrm{J}$ decreases. Therefore using the $q$-port mitigates the conflict between driving effectively the $u-g$ transition and keeping $\ket{u}$ sufficiently uncoupled to the mode. In \S\ref{sec:driving-options} we will illustrate other striking advantages of this choice when operating in the USC regime.

As in \S\ref{sec:vp-conversion}, we consider a two-tone field $W(t)$ quasi-resonant to the two relevant transitions $\omega_p = E_0 - E_{0u} - \delta_p$ and $\omega_s = E_0 - E_{2u} - \delta_s$ where $\delta_{p/s}$ are the detunings.
Insight into the problem is gained if we project $H_\mathrm{C}$ onto the subspace $\mathrm{span}\{\ket{\Psi_{0}},\ket{\Psi_{2n\,u}},\, n=0,1, \dots\}$. Treating the drives in the RW approximation and retaining one- and two-photon quasi-resonant terms we obtain the following control Hamiltonian in the rotating frame
\begin{equation}
  \label{eq:HC-RWA2}
  \begin{aligned}
    \tilde H_\mathrm{C}^\Lambda(t) \to \frac{1}{ 2} \sum_{m=0}^\infty \big[ & \Omega_{pm}(t)  \,  \ketbra{\Psi_{2m\,u}}{\Psi_0}
    \\&
      +  \Omega_{sm}(t) \, \ketbra{\Psi_{2(m+1)\,u}}{\Psi_0} \big] + \mbox{h.c.}
  \; ,
  \end{aligned}
\end{equation}
where $\Omega_{pm}(t) := \mathscr{W}_p(t) \,\braket{\Psi_{mu}}{\hat q | \Psi_0}$ and $\Omega_{sm}(t) := \mathscr{W}_s(t) \,\braket{\Psi_{mu}}{\hat q | \Psi_0}$, providing a generalization of Eq.(\ref{eq:control-lambda}). Thus the relevant part of the two-tone drive implements a chain of $\Lambda$ (and Vee) configurations $\ket{\Psi_{2m\,u}} \leftrightarrow \ket{\Psi_0} \leftrightarrow \ket{\Psi_{2(m+1)\,u}}$. By construction, the $m=0$ is the main one while the $m>0$ ones are important only if the external fields are near to the two-photon resonance, $\delta_{sm} \approx \delta_{pm}$. Thus the key quantity is the Stokes Rabi amplitude $\Omega_{s0} \propto \braket{\Psi_{2u}}{\hat q |\Psi_0}$ which must be nonzero to achieve population transfer.

Going back to the full-driven Hamiltonian we now prove that if $\ket{u}$ is uncoupled then photon-pairs produced by population transfer are only due to the faithful conversion of VPs also in the EQR model providing a test of matter-light entanglement in the ground-state $\ket{\Psi_0}$. Indeed matrix elements relevant to the protocol are
\begin{equation}
  \label{eq:dipolematrixel}
  \braket{nu}{\hat q |\Psi_0} =
  \frac{1+(-1)^n}{ 2}
  \sum_{j=g,f,\dots} q_{u,j} \,
  \braket{n,j}{\Psi_0}
\end{equation}
and in particular
$\Omega_{s0}$ is proportional to the $n=2$ amplitude showing that population transfer $\ket{0u}\to\ket{2u}$ takes place only if $\ket{\Psi_0}$ contains $n=2$ VPs, QED.
Eqs.(\ref{eq:HC-RWA2},\ref{eq:dipolematrixel}) also suggests that in the USC regime many pairs of ground-state VPs can be converted to real photons by coherent transitions in multipod linkage configurations~\cite{ka:20-wangkais-fphy-qudits}. Again, this happens {\em only if} $\ket{\Psi_0}$ has non-vanishing components containing $n$ pairs of VPs.

If the mode also couples to  the $u-g$ transition the above statements are slightly weakened. In fact, many more amplitudes contribute to the relevant matrix elements. In particular, the main Stokes amplitude has the structure
\begin{equation}
  \label{eq:stokes-amplitude}
  \braket{\Psi_{2u}}{\hat q |\Psi_0} = \sum_{nij=0}^\infty\!\!\rule[5pt]{0pt}{5pt}^\prime
  \braket{\Psi_{2u}}{ni}
  q_{ij} \braket{n j}{\Psi_0} \; ,
\end{equation}
where the prime means that the sum is restricted to even $n+i$ and odd $n+j$. Differently than before, some of the amplitudes in the sum do not vanish for
purely corotating interaction. We group
them in a quantity we denote as the RW amplitude
\begin{equation}
  \begin{aligned}
     & \braket{\Psi_{2u}}{\hat q |\Psi_0}_\mathrm{RW} :=
    \\&\qquad=
    q_{gu}  \braket{\Psi_{2u}}{1g}
    \braket{1u}{\Psi_0}+ q_{eg}  \braket{\Psi_{2u}}{0e}
    \braket{0g}{\Psi_0} \, .
  \end{aligned}
  \label{eq:RWpart-matrixel}
\end{equation}
Then the RW amplitude contains the terms of the Stokes amplitude Eq.(\ref{eq:stokes-amplitude}) "surviving" when the counter-rotating part in the interaction is switched off, i.e.
\begin{equation}
  \begin{aligned}
    \braket{\Psi_{2u}}{\hat q |\Psi_0}_\mathrm{RW} & \to
    q_{gu}  \braket{\psi_{2u}}{1g}
    \braket{1u}{\psi_0}
    \\&+ q_{eg}  \braket{\psi_{2u}}{0e}
    \braket{0g}{\psi_0} \stackrel{!}{=}
    \braket{\psi_{2u}}{\hat q |\psi_0} \, ,
  \end{aligned}
  \label{eq:RWpart-matrixel2}
\end{equation}
where $\ket{\psi_m}$ are the eigenstates of the extended JC model obtained by canceling
counter-rotating terms.
\begin{figure}[t!]
  \centering
  \includegraphics{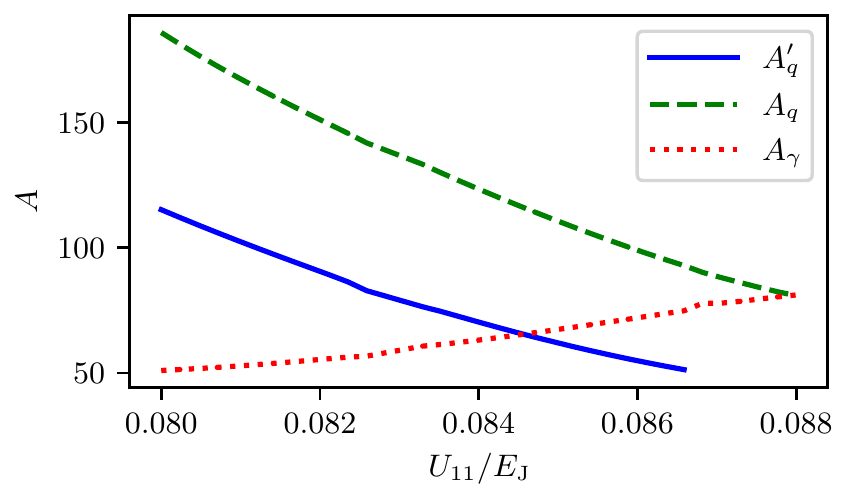}
  \caption{The figures of merit
  $A(q),\,A^\prime(q), A(\gamma)$ quantifying faithful
  detection of VPs in the EQR model Eq.(\ref{eq:Hqc}) as a function of  $U_{11}/E_\mathrm{J}$ for $E_\mathrm{J}/E_{C_1} = 0.9$ and resonant interaction with $g/\omega_c = 0.5$.
  Notice that in the region of interest, $A(q)>A(\gamma)$ indicates that the $q$-drive is more faithful than the $\gamma$-drive.
  \label{fig:figure-merit}}
\end{figure}
Now, if the RW amplitude Eq.(\ref{eq:RWpart-matrixel}) is non-zero, population transfer $\ket{\Psi_{0u}} \to \ket{\Psi_{2n\,u}}$ may in principle occur with no need for counter-rotating interaction. Nevertheless,
if the RW amplitude is sufficiently small
faithful conversion by population transfer of ground-state VP can be unambiguously guaranteed. The proof of this statement is provided in the next section. Here we support it bya physical argument. The key point is that coherent amplification is achieved {\em only if} the Rabi amplitude $\Omega_{s0}$ is larger than a non-zero (soft) threshold value.
We illustrate this fact for STIRAP operated by using in $W(t)$ pulses of width $T$ shined in the "counterintuitive" sequence~\cite{kr:17-vitanovbergmann-rmp}, i.e. $\Omega_{p/s}(t)`= F[(t\mp \tau)/T]$ with $\tau>0$. Coherent population transfer occurs {\em only if} the "global adiabaticity" condition $\max_t[\Omega_s(t)] T \gtrsim 10$ is met. Therefore if we can select values of $T$ such that
\begin{equation}
  \label{eq:T-window}
  \mathscr{W}_s^{max} T \braket{\Psi_{2u}}{\hat q |\Psi_0}_\mathrm{RW} < 10 < \mathscr{W}_s^{max} T \braket{\Psi_{2u}}{\hat q |\Psi_0}           \end{equation}
then transfer occurs via the USC channel {\em but not via the RW channel} and detected photon pairs are definitely converted VPs. A necessary condition for faithful conversion is expressed by the figure of merit
\begin{equation}
  \label{eq:figmerit-Aq}
  A(\hat q) := \frac{\braket{\Psi_{2u}}{\hat q |\Psi_0} }{ \braket{\Psi_{2u}}{\hat q |\Psi_0}_\mathrm{RW}} >10 \; ,
\end{equation}
which sets a faithfulness criterion
more rigorous than the estimate Eq.(\ref{eq:figure-merit-simple}). Alternatively, we could compare
the Stokes amplitudes in the EQR and the extended JC model and determine the criterion
\begin{equation}
  \label{eq:figmerit-Apq}
  A^\prime(\hat q) = \frac{\braket{\Psi_{2u}}{\hat q |\Psi_0} }{ \braket{\psi_{2u}}{\hat q |\psi_0}} > 10 \; ,
\end{equation}
Since we expect $A^\prime(\hat q) < A (\hat q)$ as it is the case for the relevant region of parameters (see fig.~\ref{fig:figure-merit})
asking that photon production is negligible for the extended JC model is a sufficient condition for faithful VP conversion.
\begin{figure}[t!]
  \centering
  \includegraphics{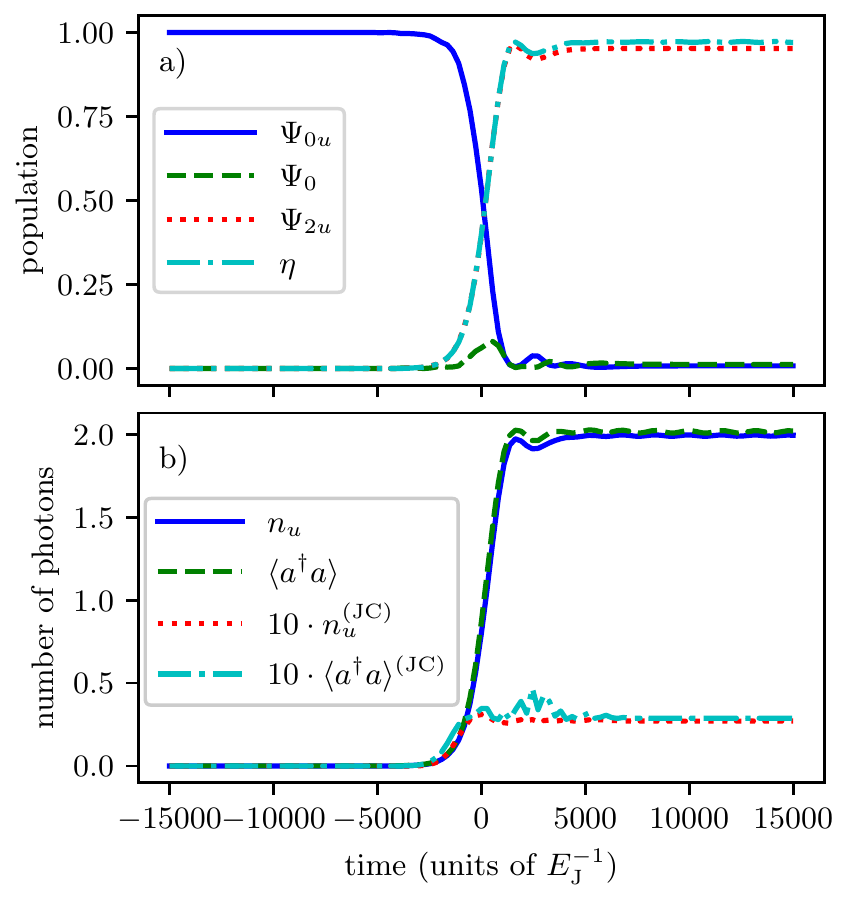}
  \caption{\label{fig:dynamics1}STIRAP dynamics at resonance for the set 2 of parameters in table~\ref{tab:circuit-parameters}, with $\Omega_0T=15$ and $T= 3000/E_\mathrm{J}$ ($T= 48\,$ns for $E_\mathrm{J}/(2 \pi)=10\,$GHz). (a) Population histories of the three states mainly involved in the protocol ($\langle \Psi_{2u}|\rho (\tf)|\Psi_{2u}\rangle \approx 0.95$) and efficiency ($\eta(\tf)\approx 0.97$) for the EQR model. (b) Evolution of the number of photons in the mode $\langle \hat n(t) \rangle$ and $n_u(t)$. At the final time for the EQR model $n_u\approx 1.99$ and $\avg{a^\dag a}\approx 2.02$ are approximately equal thus the protocol is selective. For the extended JC model, $n_u^\mathrm{(JC)}\approx \avg{a^\dag a}^\mathrm{(JC)}\approx 0.03$ are very small showing that VPs conversion is faithful.}
\end{figure}

\section{Dynamics in selected cases}
\label{sec:dynamics}
In this section, we present the central result of our work namely that the design we propose allows the efficient, faithful and selective conversion of ground-state VPs into real ones, 
which can be detected. To this end, we study the dynamics of the density matrix $\rho(t)= U(t) \, \rho(0) \, U^\dagger(t)$ of the driven system comparing pair production for the EQR and for the extended JC models. The absence of output photons for this latter implies that for the corresponding Rabi model conversion of VPs is faithful i.e. that  all the output photons are converted ground-state VPs (see Fig.~\ref{fig:figure-merit}).

We first study a STIRAP protocol operated by a two-tone drive $W(t)$ resonant with both the transitions of interest $\delta_s=\delta_p=0$. We use Gaussian pulse shapes $F(x) = \Omega_0 \,\mathrm{e}^{-x^2}$ and a delay $\tau=0.7 \,T$. We choose the amplitudes $\mathscr{W}_{p/s}(t)$ such that the resulting carrier pump and Stokes peak Rabi frequencies $\max_t[\Omega_{p2}(t)] =\max_t[\Omega_{s2}(t)] =\Omega_0$ are equal this condition guaranteeing robustness of  STIRAP~\cite{kr:01-vitanov-advatmolopt}.
We show in Fig.~\ref{fig:dynamics1}a that STIRAP successfully operates in the USC regime yielding an almost complete population transfer $\braket{\Psi_{2u}}{\rho (t_f)| \Psi_{2u}}$ while for the extended JC model practically no population transfer occurs (see Fig.~\ref{fig:dynamics1}b).
Since the transfer via multipod configurations also yields conversion of VPs we define the probability of the system to be found in the "target" subspace,  $\mathrm{span}\{\ket{nu},\, n>0 \}$
\begin{equation}
  \label{eq:efficiency}
  \eta(t) := \sum_{n>0} \mathrm{Tr}\big[ \rho(t)\,  \ketbra{n u}{nu} \big]
\end{equation}
the final value $\eta(t_f)$ being the transfer efficiency. Figs.~\ref{fig:dynamics1}ab show that for the design we propose VPs conversion has almost unit efficiency and $\sim 99\%$ faithfulness.
Another important quantity is the number of photons injected into the target subspace
\begin{equation}
  \label{eq:nu}
  n_u
  := \mathrm{Tr}\big[ \rho(t)\,\hat n \otimes  \ketbra{u}{u} \big] \; .
\end{equation}
Fig.~\ref{fig:dynamics1}b shows that practically $n_u(t)=0$ for the extended JC model confirming that for the Rabi model
$n_u(t)$ is just the number of converted VPs, i.e. population transfer is faithful.
Moreover, since  $n_u(t)$ approximately coincides with the total number of photons $\langle \hat n(t)\rangle$ almost all the converted VPs are injected into the mode leaving the AA unexcited. Leakage from the target subspace is negligibly small, thus STIRAP proves to be also highly selective.

\begin{figure}[t!]
  \centering
  \includegraphics{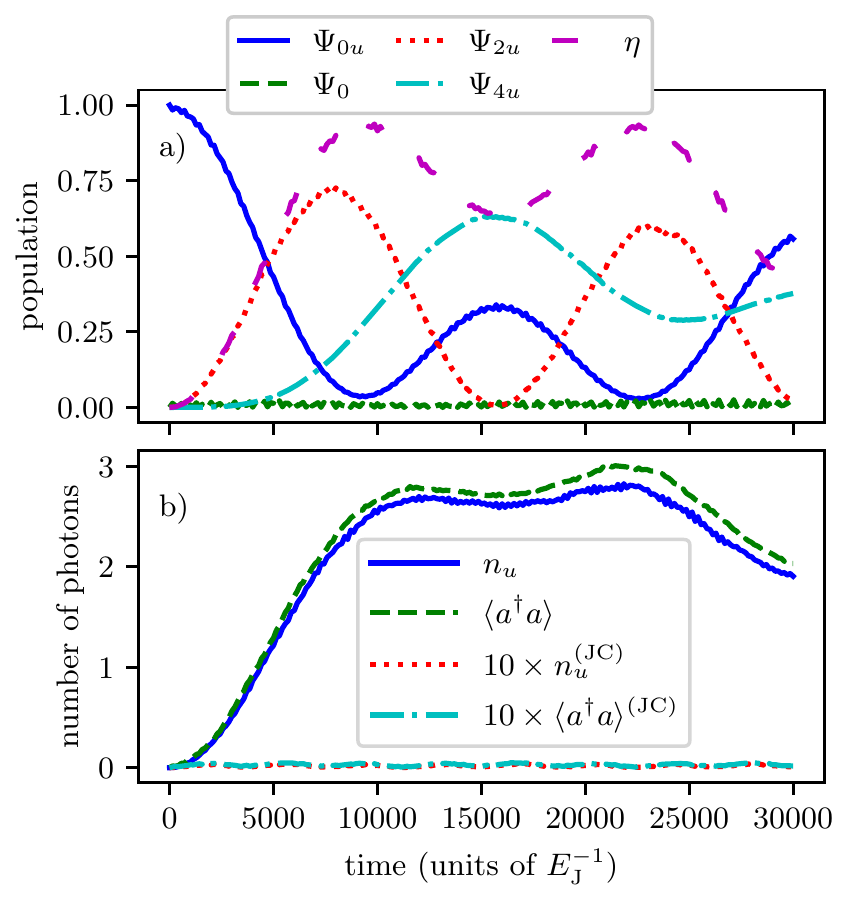}
  \caption{Raman oscillations at resonance for the same parameters of
    Fig.~\ref{fig:dynamics1} and detunings $\delta_p=\delta_s= 5 \Omega_0$. (a)
    Population histories for the EQR model. (b) Dynamics of the number of photons in the mode.
    \label{fig:dynamics2}}
\end{figure}

\begin{figure}[t!]
  \centering
  \includegraphics{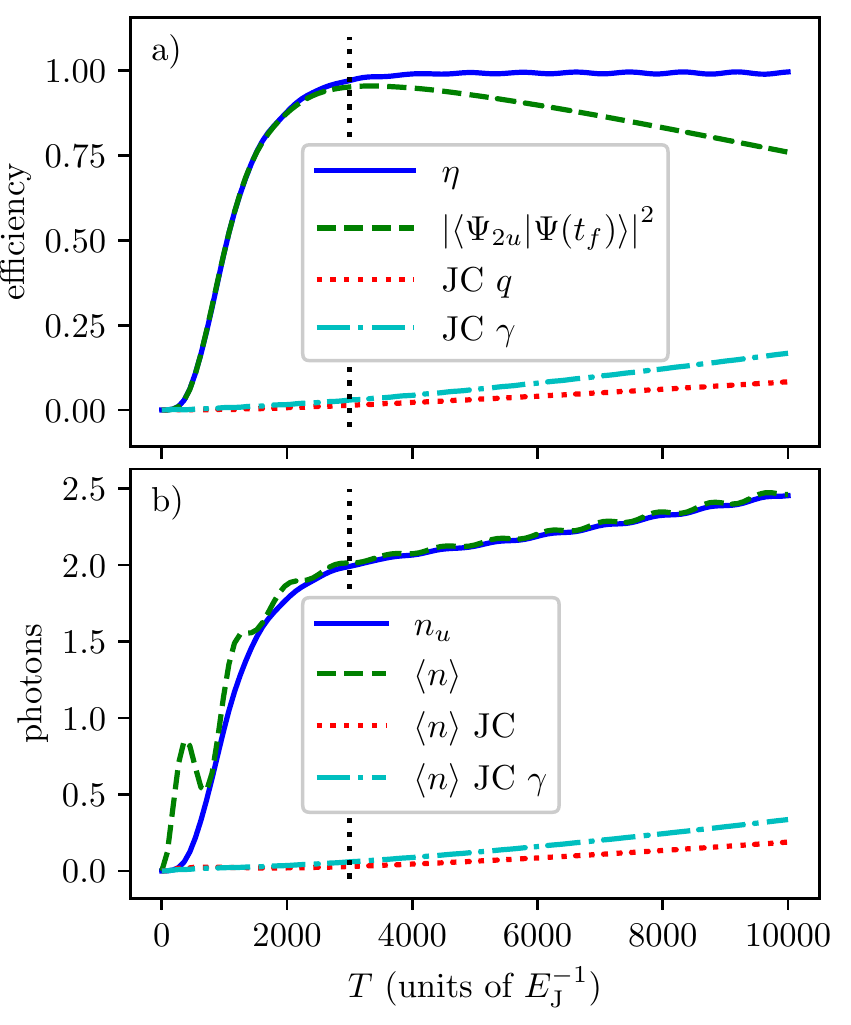}
  \caption{\label{fig:compare-full} Dynamical figures of merit for VPs conversion as a function of the time scale $T$ for a quantum circuit with parameters of set 2) in table~\ref{tab:circuit-parameters} and for fixed $\Omega_0=0.005 \, E_\mathrm{J}$. The system is driven both via the $q$-port Eq.(\ref{eq:control-q}) and the $\gamma$-port (see \S\ref{sec:driving-options}). (a) The efficiency
    $\eta(t_f)$ for the EQR model (by construction equal for the $q$- and $\gamma$-ports) and for the extended JC model. It shows that an interval of $T$ exists such that the stray interaction is ineffective the $q$-port being more faithful than the $\gamma$-port. The vertical dotted line marks the value $\Omega_{s}T=15$ where global adiabaticity is fully attained. For $E_\mathrm{J}/(2 \pi) = 10\,$GHz it requires $T = 48 \,$ns at $g/\omega_c =0.5$. Comparison with the
    population of $\ket{\Psi_{2u}}$ quantifies the impact of multipod
    transitions in the Rabi model. (b) number of photons $n_u$ transferred leaving the AA in $\ket{u}$,
    Eq.~\eqref{eq:nu}. Comparison with $\langle n(t_f)\rangle$ for EQR and JC models shows that adiabaticity guarantees a faithful and selective conversion of VPs.}
\end{figure}
Alternatively, using two largely detuned fields in the $\Lambda$ configuration with delay $\tau=0$ induces Raman oscillations between the states $\ket{\Psi_{0u}} \leftrightarrow \ket{\Psi_{2u}}$~\cite{ka:14-huanglaw-pra-uscraman}. Fig.~\ref{fig:dynamics2}b shows that again VP conversion is unambiguous. For the same parameters used in Fig.~\ref{fig:dynamics2}a the dynamics involve more states $\ket{\Psi_{2n\,u}}$. Raman oscillations ensure extremely good faithfulness and selectivity but an efficiency smaller than STIRAP requiring moreover slightly larger times.
\begin{table*}[t!]
  \centering
  \begin{tabular}{|c|c|c|c|c|c|c|c|c|c|l|l|l|}
    \hline             &
    $L\,$(nH)          & $L_2\,$(nH)            & $C_2\,$(fF)                   &
    $Z_2\,$(k$\Omega$) & $\epsilon_{eg}\,$(GHz) & $\langle \hat n \rangle_\mathrm{th}$ & $p_2^\mathrm{th}$ & $\langle n \rangle_m$  &
     $\langle n \rangle_m/ \langle n \rangle_\mathrm{th}$ 
    \\
    \hline 2           & 250                    & 955                           & 6.4        &
    12                 & 2.0                    & 0.17                          & 1.8 \%     & 0.363 & 2.13
    \\\hline 3  & 259 & 757 & 6.6 &
    10.7               & 2.2                    & 0.14                          & 1.2 \%   &    0.365 & 2.60
    \\\hline
  \end{tabular}
  \caption{Physical characterization of the systems for the sets 2-3 of Table~\ref{tab:circuit-parameters}
  for $E_\mathrm{J}/(2 \pi) = 10\,$GHz yielding
  a Cooper-pair charging energy $E_{C_1}/(2 \pi) = 11\,$GHz corresponding to $C_1 = 7\,$fF, a Josephson inductance $L_J=16\,$nH and a critical current $I_C=20\,$nA.
  A pulse with width $T = 48\,$ns and amplitude
  $\mathscr W_s^{max} =900\,$MHz ensures an efficiency
  $\eta \approx 0.97$ of VPs conversion by STIRAP. Thermal populations of the mode are evaluated for 
  $\Theta_\mathrm{eff}= 50\,$mK yielding an average $\langle n \rangle_\mathrm{th}$. This is compared with the average number of VPs emitted in a cycle $\langle n \rangle_m= P_m/(\kappa \,\hbar \omega_c)$ showing that these latter can be discriminated.
  }
  \label{tab:physical-param}
\end{table*}

In Fig.~\ref{fig:compare-full}a we study the dependence on $T$ of the total efficiency $\eta(t_f)$. For the EQR model, we obtain $\sim 100 \%$ efficiency for $T$ larger than a soft threshold set by the \textquote{global adiabaticity} condition for STIRAP, $\max_t[\Omega_s(t)] T \gtrsim 10$ \cite{kr:17-vitanovbergmann-rmp}.  We also consider here both the $q$- and the $\gamma$-port comparing their performances. For both, we use a resonant $\mathscr{W}_{p/s}(t)$ such that the maximum $\Omega_{s2}=\Omega_{p2}= 0.005\,  E_\mathrm{J}$ is the same and find that in these conditions the two curves coincide.
The same analysis for the extended JC model is carried out by using a control field with the same $\mathscr{W}_{p/s}(t)$ as before but adjusting the frequencies at resonance. Population transfer is negligible showing that VP conversion is almost perfectly faithful for a large window of values of $T$ according to the argument leading to Eq.(\ref{eq:T-window}).
Notice that the non-USC population transfer is larger for the $\gamma$-port than for the $q$-port confirming the expectation that this latter is more faithful in converting VPs. In the same figure, we also plot
$\braket{\Psi_{2u}}{\rho(t_f)|\Psi_{2u}}$ for the Rabi model showing that when adiabaticity increases multipod transitions set on.
Finally, Fig.~\ref{fig:compare-full}b shows that when the "global adiabaticity" condition is met practically all the output photons at the end of the protocol are converted VPs injected in the mode and not in atomic excitations thus the protocol selectively addresses the correct target subspace. This property holds true for sets 1-3 of Table~\ref{tab:circuit-parameters}. Summing up Figs.~\ref{fig:dynamics1}-\ref{fig:compare-full} show that we provided the sought example of a properly designed superconducting quantum circuit allowing to detect VPs by efficient, faithful and selective conversion of VPs into real photons.

\section{Implementation}
\label{sec:remarks}
\subsection{Control initialization and target state}
\label{sec:remarks-control}
Assuming a scale $E_\mathrm{J}/(2 \pi) = 10\,$GHz population transfer in Fig.~\ref{fig:dynamics1} requires a pulse width $T= 48\,$ns, the whole protocol taking a minimum time $t_T \sim 6 T \approx 300 \,$ns. We are using $\Omega_0 T =15$ thus $\Omega_0= 0.005 \,E_\mathrm{J}$ corresponds to a peak amplitude of the Stokes pulse $\mathscr{W}_s^{max}= \Omega_0/\braket{\Psi_{2u}}{\Psi_0}$  i.e.
$\mathscr{W}_s^{max}/(2 \pi)=0.09\,E_\mathrm{J}/(2 \pi) = 900\,$MHz  
for $g/\omega_c=0.5$. The same pulse amplitude $\mathscr{W}_s^{max}$ is used in Fig.~\ref{fig:compare-full} for variable pulse width  $T$. An important property of the set of parameters we found is that the target state $\ket{\Psi_{2u}}$ has a large overlap with the "uncoupled" $\ket{2u}$ making initialization and photodetection relatively simple. In particular, for samples 1-3 in Table\ref{tab:circuit-parameters} the final population $|\braket{\Psi_{2u}}{\Psi(t_f)}|^2$ is approximately the probability that a pair of photons is found in the mode. Even more importantly, the AA and the mode are almost decoupled in the final state, $\ket{\Psi_{2u}} \approx \ket{2u}$, thus the photodetection is not affected by the complications of the USC regime. Similarly, since $\ket{\Psi_{0u}} \approx \ket{0u}$ the initial state is
faithfully prepared by simply letting the system relax.

\subsection{Quantum circuit model and implementation}
\label{sec:remark-implementation}
The characteristic figures for the quantum circuit are reported in Tab.~\ref{tab:physical-param} for sets 2 and 3 of Tab~\ref{tab:circuit-parameters}.
For the latter, the coupling inductance is $L= 259\,$nH and for the mode $C_2 =6.6\,$fF and $L_2=757\,$nH yielding a cavity impedance $Z_2 = \sqrt{L_2/C_2} = 10.7\, \mathrm{k\Omega}$ of the order of the resistance quantum $R_K$.

These circuit elements can be implemented by super-inductor technology employing high kinetic inductance films or Josephson junction arrays~\cite{ka:09-manucharyan-science-fluxonium,ka:12-maslukpopdevoret-prl-superinductor,ka:19-gruenhaupt-natmaterials-superinductance,ka:19-hazard-prl-superinductances}, and in particular impedances of several k$\Omega$ have been recently obtained~\cite{ka:20-peruzzofink-prapplied-superinductance,ka:19-zhanggershenson-prappl-superinductor}. The AA design is reminiscent of a fluxonium and its fabrication is within reach for present technologies. We exploit however unconventional features such as the symmetric bias leading to a peculiar multilevel structure and the "light-mass" $C_1$ avoiding trapping the AA in the side minima of the potential.
This is crucial to guarantee large enough dipole matrix elements for both the coupling ($\gamma_{ge}$) and the control ($\gamma_{ug}$). On the contrary, in the standard fluxonium qubit, the design favours trapping in one of the side minima
to suppress the relaxation rate $\propto |\gamma_{ug}|^2$.
Notice that for us the smaller $E_\mathrm{J}/E_{C_1}$ the better but this would pose problems to the implementation of the Josephson junction.

\begin{table*}[t]
  \centering
  \begin{tabular}{|c|p{45mm}|p{62mm}|p{50mm}|}
    \hline                                       &
    {\bf Requirement}                            & {\bf Problem}                                                                                                                  & {\bf Solution}                                   \\\hline 1&
    Local control on the AA                      & Avoiding stray driving of the mode                                                                                             & Driving via $q$-port,
    \S\ref{sec:driving-options}
    \\\hline 2& Efficiency for the USC channel& Inducing population transfer via the USC channel despite the small $\braket{2g}{\Psi_0}$&  Global adiabaticity Eq.(\ref{eq:T-window}) for the USC channel by large $T \mathscr{W}_s$ and $g$
    \\\hline 3& Faithful and selective conversion of VPs
                                                 & Suppressing stray processes (population transfer via the JC channel, Stark shifts, population of higher-energy AA eigenstates) &
    No adiabaticity for the JC channel Eq.(\ref{eq:T-window});
    not too large $g$ and $\mathscr{W}_s$
    \\\hline 4&
    Small dephasing $T \ll T_\phi \sim 1/\kappa$ & Dephasing of the mode reduces efficiency                                                                                       & Not too large $\kappa$,  \S\ref{sec:decoherence}
    \\\hline 5& Large detection efficiency & Discriminating power emitted by oscillator decay from thermal floor & Large enough $\kappa\, \to$
    tradeoff design of measurement
    \S\ref{sec:decoherence}
    \\\hline 6&
    Faithful preparation/detection               & Nearly uncoupled initial and final state                                                                                       &
    Design tradeoff, not too large $g$, large anharmonicity \S\ref{sec:decoherence}
    \\\hline
  \end{tabular}
  \caption{Requirements for faithful  control and measurement for ground-state VPs detection.}
  \label{tab:requirements-dynamics}
\end{table*}

\subsection{Decoherence and measurement}
\label{sec:decoherence}
The main features of USC are in general robust against dissipation and
in particular, the USC ground state still contains VPs~\cite{ka:17-deliberato-natcomms-uscdiss}. In our case,
the eigenstates $\{\ket{i},\,i=1,2,3\}$ of the AA are delocalized in the "physical" $\hat \Phi_1$-space (see Fig.~\ref{fig:artificial-atom}) thus they are rather insensitive to flux noise whereas charge noise is limited by the relatively small matrix elements $q_{ij}$ between the excited states of the AA participating to the EQR model.

On the other hand, noise affects coherent population transfer in both STIRAP and Raman protocols which are sensitive to decoherence in the "trapped" subspace spanned by two or more states $\{ \ket{\Psi_{2n\,u}},\,n=0,2,\dots  \}$. In our case, it is practically a low-energy subspace of the uncoupled mode, $\mathrm{span}\{\ket{n}, \, n=0,1,2\}$.
Thus the main detrimental processes are expected to be due to relaxation~\cite{ka:12-spagnolo-aphyspol-relaxation} with rate $\kappa$ in this subspace, a result which emerges from a simplified model for STIRAP~\cite{ka:21-ridolfo-epjst-virtualphotonopen} which also shows that decoherence rates $\gamma$ of the AA are not relevant. Indeed the AA ideally always sits in its ground state $\ket{u}$ and its dephasing is minimized since we operate at a symmetric point, $\gamma_x = Q_\mathrm{x} =0$.

Pure dephasing of the uncoupled mode determines a reduction in the population transferred to $\ket{\Psi_{2n\,u}}$. For 3-level STIRAP with Gaussian pulses, it can be estimated by $p_{2} = \frac{1 }{ 3} + \frac{2 }{ 3} \mathrm{exp} \big[- 3 \kappa_\phi T^2/(16 \tau)]$~\cite{ka:04-ivanov-pra-stirapdephasing} where in our case $\kappa_\phi = 3 \kappa/2$. This limits the pulse width $T$ of the driving fields and the duration of the protocol. The mode decays after starting to populate thus relaxation {\em and} dephasing are effective only in the second part of STIRAP.

For oscillators with quality factor $Q \gtrsim 10^4$ the population of the mode remains large enough~\cite{ka:21-ridolfo-epjst-virtualphotonopen} to allow photons to be detected (and even counted) by single-shot non-demolition measurements performed by a quantum probe coupled dispersively to the mode~\cite{ka:11-eichlerwallraff-prl-photontomography,ka:21-curtisschoelkopf-pra-pnrdetection}. Assuming an effective cryostat temperature $\Theta_\mathrm{eff} = 50 \,$mK the thermal population of the mode is $\langle \hat n \rangle_\mathrm{th} \ll 2$ thus the probability of detecting two thermal photons is smaller than a few per cent
(see Tab.~\ref{tab:physical-param})
allowing converted VPs to be discriminated from the thermal floor.

An experimentally less demanding procedure is a continuous measurement~\cite{ka:18-distefanopaternostro-prb-measurethermo} of the decay of converted VPs into a transmission line. A  detectable signal is obtained if $\kappa$ is large enough. In the simplest instance, the transmission line is coupled to the mode during the whole protocol.
If the decay of $\ket{\Psi_{2u}}$ before the completion of STIRAP is neglected the number of photons emitted via the double decay $\ket{2u} \to \ket{1u} \to \ket{0u}$ is given by $n_\mathrm{out}(t_m) \approx 2 p_{2}\,\big( 1 - \mathrm{e}^{- \kappa t_m} \big)$ where $t_m$ is the duration of the measurement. Here we estimate the figures for the parameters we selected and for $\kappa = 1/(3T)$ corresponding to $Q = \omega c /\kappa = 1800$ emission of two photons to an always-on coupled transmission line requires the decay to be effective for a time $t_m \sim 4 \times 10^4 /E_\mathrm{J} = 640\,$ns.
After emitting the two photons  the system is reset to the initial state $\approx \ket{0\,u}$ and the protocol can be operated sequentially. At each repetition the energy $2 \hbar \omega_c$ is emitted which corresponds to a power $P_m\approx 2\hbar\omega_c/(3T+t_m)\approx 2\hbar\omega_c\kappa/2\pi\approx3\times 10^{-18}~\text{W}$ (see Tab. III), which can be amplified by standard HEMT circuitry and discriminated with respect to thermal noise.
The total measurement time $\tau_m\approx  100\ \mu$s needed for such discrimination can be determined from the equation
$$\frac{\delta P}{P}=\frac{P_m}{k_bT_N\kappa}=\frac{1}{\sqrt{\kappa\tau_m}}= 1 \; ,$$ where $T_N\approx 2$~K is the noise temperature of the HEMT amplifier.  In other words, our experimental procedure requires hundreds of repetitions to distinguish the signal of VPs conversion from amplifier noise which is still a reasonable number. For example, in Ref.~\onlinecite{Bozyigit2011} it was shown that  even the  detected noise power of the amplifier dominates by a factor of $\sim~700$ over the single-photon power, such power is still observable using sufficient averaging. Actually, we expect even better figures since photodetection by continuous measurement may benefit also from decay {\em during} STIRAP~\cite{23:giannellianfuso-epjst-fourlevelmeas}.

For the Raman protocol, decoherence provides the mechanism for suppressing the stray RW channel when $\Omega_s$ is small but nonzero determining a soft threshold which is less selective than the global adiabaticity condition for STIRAP. Photodetection by decay is more invasive than in STIRAP where the tradeoff between relaxation and decoherence during the protocol and efficient detection is more favourable.

Relaxation of the  AA does not affect the ideal protocols where only states $\ket{2n\,u}$ are populated but it helps in non-ideal cases when it provides a further mechanism to reset the system. Dephasing of the AA is not relevant for our protocols and in any case, it is minimized by operating at a symmetric point, $\gamma_x = Q_\mathrm{x} =0$.

Finally, we observe that more elaborated measurement schemes allow faithful detection in "borderline" regions of the space of parameters.
For instance, if extra photons are produced by climbing the JC ladder -- as for sets 4-6 in Table~\ref{tab:circuit-parameters} --
VPs can be discriminated by post-selection after measuring the qubit.

\subsection{Driving options and control}
\label{sec:driving-options}
The AA could also be driven via the $\gamma$-port, i.e. by modulating external fluxes $\Phi_{xi}(t)$ which couple to the AA phase $\hat \gamma$. The control is described by $H_\mathrm{C}(t) = W_\gamma(t) \,\hat \gamma$ and can be analyzed along the same lines of \S\ref{sec:connection-dyn} since Eq.(\ref{eq:q-gamma}) implies that $\hat q$ and $\hat \gamma$ have the same selection rules.

Although driving via the $\gamma$-port is the main option in many qubits implemented by superconducting devices in our case it brings important disadvantages. Quantitative differences are apparent in Fig.~\ref{fig:compare-full} showing that using the $q$-port for driving yields a more faithful conversion of VPs. Therefore this latter choice may mitigate the conflict between requirements 4-6 of Tab.~\ref{tab:requirements-spectral}, as we argued in \S\ref{sec:connection-dyn}, and as it is also quantified by the different values of the figures of merit $A(\hat q)>A(\hat \gamma)$ for the samples 1-4 in Table~\ref{tab:circuit-parameters}).

 The other decisive advantage of the $q$-port is that the voltage drive implements naturally a {\em local control of the AA}. On the contrary using time-dependent magnetic fluxes $\tilde{\Phi}_{xi}(t)$ in the  Hamiltonian Eq.(\ref{eq:Hqc}) would require a gradiometric configuration to avoid stray direct driving of the mode which becomes a large effect in the USC regime (see Appendix~\ref{sec:control}).

It is interesting to foresee the effect of large driving fields which in principle may improve adiabaticity and/or speed up the protocol reducing the impact of decoherence and increasing the power emitted. Notice that a large $\mathscr{W}_s(t)$ is a mandatory if $g$ is not too large to obtain large enough $\Omega_{s2}$ with small $\braket{2g}{\Psi_0}$. However, large $\mathscr{W}_{s/p}$ have a negative impact on the efficiency~\cite{ka:19-falci-scirep-usc} since the external fields coupled to all the atomic transitions induce induced Stark shifts affecting the two-photon resonance condition $\delta_p(t)=\delta_s(t)$ and deteriorating coherent population transfer. Such errors are "correctable" by operating with tailored chirped-frequency or three-tone control fields~\cite{ka:16-distefano-pra-twoplusone,ka:19-falci-scirep-usc}. In the present work, the atom-mode interaction is large enough to minimize such errors by keeping the field amplitudes small. Optimal control is likely to yield a much faster protocol with larger $\mathscr{W}_{s/p}$. Large $\mathscr{W}_{s/p}$ may also produce uncorrectable errors, as population transfer via the RW channel or unwanted transitions in the multilevel structure, poisoning the output.

\section{Conclusions}
\label{sec:conclusions}
Since the early days of research on USC, detecting VPs in the entangled ground state of a quantum system has been a grail which has been progressively buried under the experimental challenges it poses.
In this work, we propose a solution to this long-standing problem that leverages state-of-the-art quantum technologies. We show how experimental challenges can be overcome, which is possible though not immediate. This is perhaps why, despite the rich physical scenario offered by the USC regime experiments so far have been limited to spectroscopy. Our proposal combines various advanced ingredients. The first one is the design of an unconventional superconducting multilevel AA reminiscent of a  fluxonium~\cite{ka:09-manucharyan-science-fluxonium} qudit but flux-biased at an unusual symmetry point, designed such as to have a \textquote{light mass}, and controlled by voltages and not by magnetic fluxes. Second, the galvanic coupling to an electric resonator is implemented by last-generation superinductors~\cite{ka:20-peruzzofink-prapplied-superinductance,ka:19-zhanggershenson-prappl-superinductor}. Third, the output signal of the detected ground-state VPs is coherently amplified using advanced control whose proper tailoring provides efficient, faithful and selective conversion of VPs to real ones. Finally, we propose a simple continuous measurement protocol of the output photons calibrated to achieve a positive tradeoff between decoherence and detection efficiency. Both STIRAP and Raman oscillation can trigger the coherent conversion of VPs the former technique being preferable for its robustness
and remarkable resilience to the backaction of the measurement. The implementation of the whole experimental setup is feasible with present-day superconducting quantum technologies~\cite{kr:20-kjaergaardoliver-annurevcond-revsupqubits}.

Our proposal fulfils the stringent requirements for the three-level detection technique. Anharmonicity of the AA spectrum is of paramount importance since mixing the oscillator states of the EQR model yields energy spectra that are far from what is desired. Another key point is the availability of  AA "ports"  providing at the same time USC with the mode and faithful VPs conversion. Finally, the design we selected implements an intrinsic switching mechanism of USC allowing preparation and measurement in states where the AA and the mode are effectively decoupled.

These conditions are met in a narrow region of the space of parameters, which suggests an explanation for the lack of experiments. Remarkably this region lies in the so-called "non-perturbative USC regime"~\cite{kr:19-forndiazsolano-rmp-usc}, $g\sim 0.5$, ensuring that truly non-perturbative physics can be observed.

The space of the parameters could be enlarged by optimizing both design and protocol. Optimal design could be systematically searched for by using the recipe of \S\ref{sec:case-study}.
Advanced computational methods of data analysis~\cite{ka:21-brown-njp-rlcoherentpop,ka:22-giannelli-pla-tutorial} could extend the investigation including other parameters, such as the static bias.

Faster protocols may be found by optimal control theory~\cite{koch2022quantum} also exploiting multipod transitions and integrated measurement. Measurement schemes with post-selection could allow the handling of multilevel systems with more complicated spectra.

Some of the requirements for USC-selective population transfer can be softened thus enlarging the space of parameters and platforms where an experiment may be successful. For instance, some transient population of the intermediate state is tolerable, softening the adiabaticity requirement as well as decoherence times $T_\phi \sim T$ and the production of a small number of RW-photons.

We finally observe that a successful experiment would also be the first direct demonstration of coherent dynamics in the USC regime. Since the dynamics is adiabatic and involves only the USC ground state its demonstration is less demanding than for coherent dynamics in the USC manifold. Therefore it could be a benchmark for quantum control at USC, paving the way for appealing applications to quantum technologies.

\acknowledgments
We acknowledge F. Deppe, V. Villari, G. Chiatto and G. Anfuso who helped to develop key insights for this paper. We acknowledge discussions and remarks by  S. De Liberato, S. Felicetti, P. Forn-Diaz, N. Roch, J. Ankerold, J. Koch, 
A. Ridolfo, and W.W. Huang.
\\
This work was supported by the QuantERA grant SiUCs (Grant No. 731473), by the PNRR MUR project PE0000023-NQSTI, by ICSC – Centro Nazionale di Ricerca in High-Performance Computing, Big Data and Quantum Computing,
by the University of Catania, Piano Incentivi Ricerca di Ateneo 2020-22, project Q-ICT. EP acknowledges the COST Action CA 21144 superqumap and
GSP acknowledges financial support from the Academy of Finland under the Finnish
Center of Excellence in Quantum Technology QTF (projects 312296, 336810, 352927,
352925).

\appendix

\section{Galvanically coupled quantum circuits}
\label{sec:qcircuits}
We consider the circuit in Fig.2 of the main text consisting in two superconducting loops denoted by $i=1,2$. Each loop contains a capacitance $C_i$ with associated flux variable $\Phi_i$.
Loop 1 contains also a Josephson junction whose state variable is $\Phi_1= \hbar \gamma/(2e)$
where $\gamma$ is the junction's gauge-invariant phase. It is connected to the ground via two equal capacitances $C_\mathrm{g}$, with flux variables $\Phi_\mathrm{gL}$ and $\Phi_\mathrm{gR}$, and the voltage source $V_g$.
We define a flux variable relative to the inductive elements in each loop $\Phi_{Ii}$ and the flux bias parameters $\Phi_{xi}$ relative to the external magnetic fields concatenated with the loops. These quantities obey the loop constraints
\begin{equation}
  \label{eq:flux-constraints}
  \left\{
  \begin{array}{ll}
    -\dot{\Phi_{gL}}+\dot{\Phi}_1 + \dot{\Phi_{gR}} + V_g =0 &
    \\
    \Phi_i + \Phi_{xi} + \Phi_{Ii} = 0                       & \quad i=1,2
  \end{array}
  \right. \; .
\end{equation}
We choose as Lagrangian coordinates the two $\Phi_i$s and $\Phi_g  :=(\Phi_{gL}+\Phi_{gR})/2$ and write
the Lagrangian
\begin{equation}
  \label{eq:lagrangian}
  \begin{aligned}
    {\cal L}_T =  \frac{C_1 + C_g/2 }{ 2}
    \, \dot \Phi_1^2 - Q_\mathrm{x}\, \dot \Phi_1
    + \frac{C_2 }{ 2} \, \dot \Phi_2^2 +
    C_g \, \dot \Phi^{2}_g
    \\
    + E_\mathrm{J} \,\cos \frac{2 \pi \Phi_1 }{ \Phi_0} - U_L(\Phi_i|\Phi_{xi})
   \; . \end{aligned}
\end{equation}
The kinetic term is the electrostatic energy which depends on the bias charge parameter $Q_\mathrm{x}={C_g V_g/2}$.
The last potential term is found from the inductive energy
$U_L = \frac{1 }{ 2} \sum_{ij} I_i \, \mathbbm{L}_{ij} \, I_j$ where $\mathbbm{L}$ is the inductance matrix
defined by the relation
$\Phi_{Ii} = \sum_{ij=1}^2 \mathbbm{L}_{ij} \, I_j$ where counterclockwise currents in the loops are taken positive.
For the circuit in Fig.2 of the main text  we obtain
$$
  \mathbbm{L} =
  \left(\hskip-3pt
  \begin{array}{cc}
      L_1 + L & - L + M
      \\
      -L + M  & L_2 + L
    \end{array}\hskip-3pt
  \right)  \; ,
$$
where $M>0$ is the mutual inductance. The inductive potential can now be expressed in terms of the Lagrangian coordinates as
$$
  \begin{aligned}
    U_L & =  \frac{1 }{ 2} \sum_{ij=1}^2 \Phi_{Ii} \, [\mathbbm{L}^{-1}]_{ij} \, \Phi_{Ij}
    \\
        & = \frac{1 }{ 2}  \sum_{ij} (\Phi_{i}+\Phi_{xi}) \, [\mathbbm{L}^{-1}]_{ij} \,  (\Phi_{j}+\Phi_{xj})
         \; .
  \end{aligned}
$$
The coordinate $\Phi_g$ is cyclic and the corresponding momentum $Q_g$ is conserved thus this degree of freedom will be hereafter ignored.

For constant external magnetic bias we can redefine $\Phi_i \to \Phi_{i} + \Phi_{xi}$  obtaining
\begin{equation}
  \label{eq:lagrangian0}
  \begin{aligned}
    {\cal L} & =  \frac{C_1 + C_g/2 }{ 2}
    \, \dot \Phi_1^2 - Q_\mathrm{x}\, \dot \Phi_1
    + \frac{C_2 }{ 2} \, \dot \Phi_2^2 +
    \\ &
    + E_\mathrm{J} \,\cos \frac{2 \pi (\Phi_1 - \Phi_{x1}) }{ \Phi_0} - \frac{1}{ 2}  \sum_{ij} \Phi_{i} \, [\mathbbm{L}^{-1}]_{ij} \,  \Phi_{j}
     \; .
  \end{aligned}
\end{equation}
Notice that the constant bias $\Phi_{x2}$ is gauged away. From the canonical momenta
$Q_1  = (C_1 + C_g/2)\, \dot \Phi_1 - Q_\mathrm{x}$ and
$Q_2 = C_2\, \dot \Phi_2$
the Hamiltonian of the circuit is found
\begin{equation}
  \label{eq:H-undriven}
  \begin{aligned}
    H_{qc} & =  \frac{(Q_1+ Q_\mathrm{x})^2 }{ 2 C_1 + C_g} +
    \frac{Q_2^2 }{ 2 C_2}
    \\ &\!\!
    - E_\mathrm{J} \,\cos \frac{2 \pi (\Phi_1 - \Phi_{x1}) }{ \Phi_0} + \frac{1}{ 2}  \sum_{ij} \Phi_{i} \, [\mathbbm{L}^{-1}]_{ij} \,  \Phi_{j}  \, ,
  \end{aligned}
\end{equation}
which is finally quantized in the canonical way yielding the quantum circuit Hamiltonian $H_{qc}$ Eq.5 of the main text). This form is convenient for the subsequent analysis since external parameters enter only the AA Hamiltonian $H_\mathrm{AA}$.

Typically, $C_1 \gg C_g$ thus this latter capacitance can be neglected in the denominator of Eq.(\ref{eq:H-undriven}). Also,
$L \gg M$ which then can be neglected in
$\mathbbm{L}$ yielding
\begin{equation}
  \label{eq:invers-ind}
  \mathbbm{L}^{-1} = \frac{1}{ L L_1 + L L_2 + L_1 L_2}
  \left(\hskip-3pt \begin{array}{cc}
      L+L_2 & L
      \\
      L     & L+L_1
    \end{array}\hskip-3pt\right)  \; .
\end{equation}
In the limit $C_g \to 0$ also the charge bias drops and the momentum $Q_1 = C_1 \dot \Phi_1$ becomes the charge at the capacitor $C_1$. Since in general both $\hat Q_i$
are related to charges on the surface of capacitors their spectrum is continuous and the wavefunctions belonging to the Hilbert space are defined for $\Phi_i \in ]-\infty,\infty[$.

\section{Control Hamiltonian}
\label{sec:control}
Control is operated by external fields introduced by adding time-dependent parts $\Phi_{xi} \to \Phi_{xi} + \tilde \Phi_{xi}(t)$ and $Q_{g} \to Q_{g} + \tilde Q_{g}(t)$ to the external parameters in the Lagrangian.  Eq.(\ref{eq:lagrangian}). Then Eq.(\ref{eq:lagrangian0}) is written by
keeping the time-dependent component
$\tilde{\Phi}_{x1}(t)$ in $U_L$.
The corresponding Hamiltonian has a time-independent part given by Eq.(\ref{eq:H-undriven}) and a control part obtained by grouping the time-dependent terms
\begin{equation}
  \label{eq:HC}
  H_\mathrm{C}(t) =  \frac{\tilde Q_\mathrm{x}(t) }{ C_1} \, \hat Q_1
  +  \sum_{i} \tilde\Phi_{xi}(t) \, [\mathbbm{L}^{-1}]_{ij} \, \hat \Phi_{j}
   \; .
\end{equation}
Modulating only $Q_\mathrm{x}(t)$ by an external voltage $V_g(t)$ yields the local control Eq.13 of the main text we use in this work.
Instead, the standard drive by external fluxes is described by the control Hamiltonian
\begin{equation}
  \label{eq:HC-flux}
  \begin{aligned}
    H_\mathrm{C}(t) = &
    \Big\{  \tilde{\Phi}_{x1}(t) \,[\mathbbm{L}^{-1}]_{11} +
    \tilde{\Phi}_{x2}(t) \,[\mathbbm{L}^{-1}]_{21}
    \Big\}
    \,\hat \Phi_1
    \\& +
    \Big\{  \tilde{\Phi}_{x1}(t) \,[\mathbbm{L}^{-1}]_{12} +
    \tilde{\Phi}_{x2}(t) \,[\mathbbm{L}^{-1}]_{22}
    \Big\}
    \,\hat \Phi_2  \; ,
  \end{aligned}
\end{equation}
so clearly $\tilde{\Phi}_{xi}(t) $ does not drive the circuit locally.
For instance, the flux $\Phi_{x1}(t)$ piercing the AA also couples to the mode's coordinate $\hat \Phi_2$ the partition ratio being
$[\mathbbm{L}^{-1}]_{12}/[\mathbbm{L}^{-1}]_{11} =L/(L+L_2)$ which in the USC regime may be of order one. A local drive could be obtained by operating with both fluxes $\Phi_{xi}(t)$ in a "gradiometric" configuration, $\tilde{\Phi}_{x2}(t) = - \tilde{\Phi}_{x1}(t) \,[\mathbbm{L}^{-1}]_{12}/[\mathbbm{L}^{-1}]_{22}$. For the case study $L_1=0$ treated in this work  $\tilde{\Phi}_{x2}(t) = - \tilde{\Phi}_{x1}(t)$ and we obtain the control Hamiltonian $H_\mathrm{C}= W(t)\,\hat \gamma$ with $W(t)= \hbar \tilde \Phi_{x1}/(2 e L)$.

Of course, in view of the complexity of the spectrum of the EQR model operating directly with local control by the external voltage is the better option. It is worth stressing that since matrix elements of the "momentum" $\hat Q_1(t)$ are more off-diagonal than those of the coordinate $\hat \Phi_1(t)$ care is required in the truncation of the model.

We remark that modelling a few-level AA requires some care since canonically equivalent Hamiltonians have different sensitivity to the truncation of the Hilbert space~\cite{ka:18-debernardisrabl-pra-cqednonperturbative} giving rise to the so-called \textquote{gauge ambiguities}~\cite{ka:19-distefanosavasta-natphys-gaugeambiguities}. Since we considered a large number of eigenstates of the AA there is no ambiguity in our case.
Notice that the equivalent of a diamagnetic term emerges naturally in $H_{qc}$  in the dependence of the diagonal entries of $\mathbbm U$ on the coupling inductance $L$ (see Appendix~\ref{sec:qcircuits}) which thus renormalizes both the bare natural frequency of the mode and the AA potential.

\section{Data analysis}
Eigenvectors of $H_\mathrm{AA}$ depend on the parameters $x:=E_\mathrm{J}/E_{C_1}$ and $y:=U_{11}/E_\mathrm{J}$
while eigenvalues depend also on the scale $E_\mathrm{J}$. We calculate the spectrum, in particular
$\epsilon_{eg}/E_\mathrm{J}$ and $\epsilon_{fe}/E_\mathrm{J}$ and the ratios
$\gamma_{ug}/\gamma_{ge}$ and $\gamma_{ef}/\gamma_{ge}$. We then fix
$g/\epsilon_{eg}$ and evaluate $A$. This yields most of the results
shown in Fig.~5 of the main text.
We then proceed by imposing the conditions
$$
  \omega_c = \sqrt{2 E_{C_2} U_{22}} \stackrel{!}{=}\epsilon_{eg} \quad ;
  \quad U_{12} \, \Big(\frac{E_{C_2} }{ 2 U_{22}}\Big)^{1/4} = \frac{g }{ \gamma_{ge}}
   \; .
$$
For $L_1 = M=0$ we have another equation $U_{12}=U_{22}$, and the three equations can be inverted yielding
\begin{equation}
  U_{22} = \frac{2 }{ \varepsilon}\, \Big(\frac{g }{ \gamma_{ge}}\Big)^2
  \quad ; \quad
  E_{C_2} = \frac{\varepsilon^3  4}\, \Big(\frac{ \gamma_{ge} }{ g}\Big)^2
   \; .
  \label{eq:parameters1}
\end{equation}
Parameters in Table~II of the main text are obtained by expressing $U_{ij}$ in terms of the inductances $(L,L_2)$
$$
  L_2 = \frac{(\Phi_0/ 2\pi)^2 }{ U_{22}}
  \quad ; \quad
  L = \frac{(\Phi_0/ 2\pi)^2 }{ U_{11} - U_{22}}
   \; .
$$
The inductance matrix is positively defined if and only if $L, L_2>0$ therefore values of
$(x,y,g)$ such that $U_{11} < U_{22}$ are not acceptable.
They correspond to the regions above the blue lines
in Fig.~5 of the main text.

\bibliography{atom-cavity,circuit-QED,USC,stirap,VPdesign,superconducting-qubits}

\end{document}